\documentclass[useAMS,usenatbib,usegraphicx]{mn2e}

\usepackage{times}
\usepackage{epsfig}
\usepackage{epstopdf}
\newcommand{\kms}{\mbox{${\rm km\;s^{-1}}$}}

\newcommand{\sauron}{{\texttt {SAURON}}}
\newcommand{\Vimos}{{\texttt {VIMOS}}}

\newcommand{\HST}{{\it HST\/}}
\newcommand{\Oiii}{[{\sc O$\,$iii}]}

\newcommand{\Ha}{H$\alpha$}
\newcommand{\Hb}{H$\beta$}

\newcommand{\Ni}{[{\sc N$\,$i}]}
\newcommand{\Nii}{[{\sc N$\,$ii}]}
\newcommand{\Sii}{[{\sc S$\,$ii}]}
\newcommand{\OiiioHb}{\Oiii/\Hb}

\newcommand{\mnras}{MNRAS}

\newcommand{\placefigApertures}{
  \begin{figure}
    \begin{center}
      \includegraphics[width=\columnwidth]{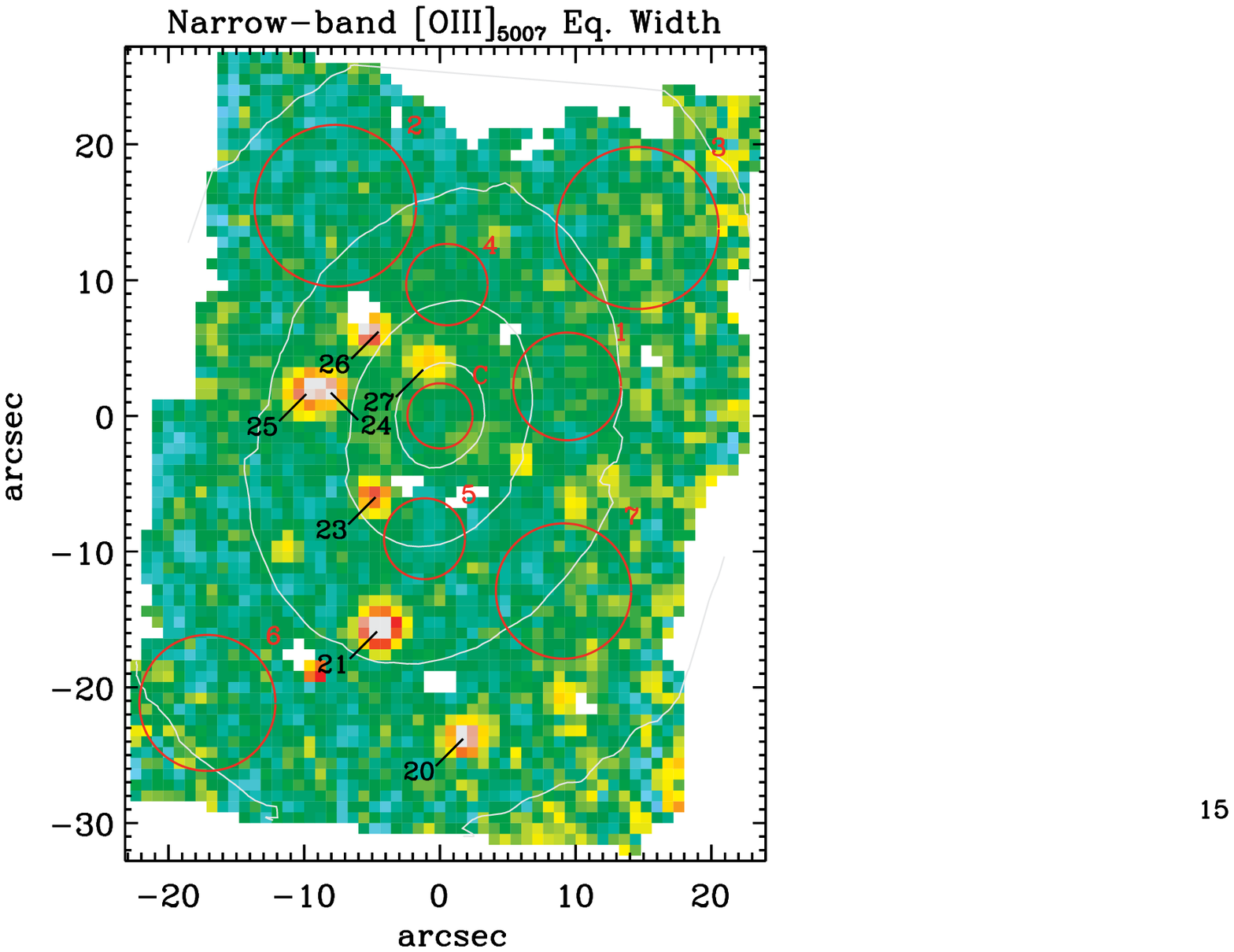}
    \end{center}
    \caption[]{Map of the equivalent width (EW) of the estimated
      \Oiii$\lambda5007$ flux from narrow-band imaging based on the
      \sauron\ data. The flux in the \Oiii\ region is measured through
      a 2.5\AA-wide spectral window centred on the expected position
      of \Oiii\ given the systemic velocity of $V_{sys}=-197\kms$ for
      M32, whereas the flux of the stellar continuum is computed using
      the mean spectral energy density across the entire spectrum
      times the same 2.5\AA\ wavelength interval. The difference
      between these two flux measurements gives an estimate of the
      \Oiii\ flux, which once divided by the mean spectral energy
      density in the stellar continuum leads to the mapped EW values.
      Isolated patches of positive EW values could originate from the
      unresolved \Oiii\ emission of PNe, and in fact the brightest and
      most circular sources correspond to the 7 PNe that were already
      identified in the optical regions of M32 by \citet[][shown and
        labelled in black]{Cia89}.
      The red circles delimit galactic regions that are most likely
      devoid of any nebular emission and where the \sauron\ spectra
      were added up to obtain spectra of very high quality. In turn,
      these high-quality spectra were used to derive the optimal
      templates that we adopted to describe the stellar continuum in
      our data (see text for more details).}
    \label{fig:Apertures}
  \end{figure}
}

\newcommand{\placefigAoN}{
\begin{figure}
\begin{center}
\includegraphics[width=\columnwidth]{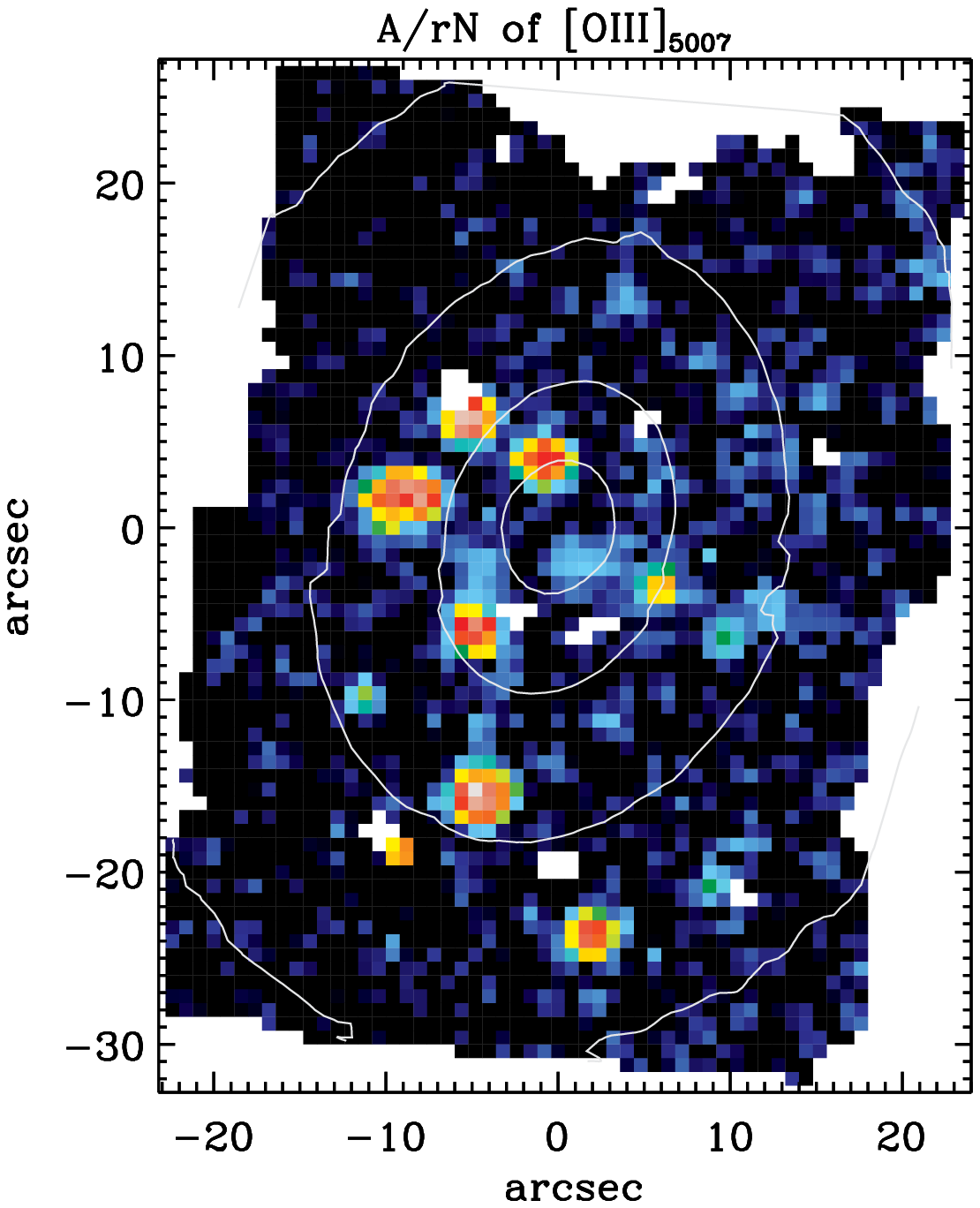}
\end{center}
\caption[]{Map of the values of the $A/rN$ ratio of the
  \Oiii$\lambda5007$ emission measured in the \sauron\ spectra. The
  $A/rN$ values are shown in a logarithmic scale and dark-blue bins
  identify regions where $A/rN < 3$ and the \Oiii\ line is not
  formally detected. Several of the weak positive patches in
  Fig.~\ref{fig:Apertures} correspond here to circular sources with
  gently declining $A/rN$ profiles as would be expected from
  atmospheric scattering of the \Oiii\ emission of an unresolved PN.
  This is not the case for the rectangular patch approximately
  20\arcsec\ to south and east of the centre, which is a spurious
  measurement where in fact no \Oiii$\lambda4959$ emission is detected
  along with the larger \Oiii$\lambda5007$ line.}
\label{fig:AoN}
\end{figure}
}

\newcommand{\placefigFluxPNe}{
\begin{figure}
\begin{center}
\includegraphics[width=\columnwidth]{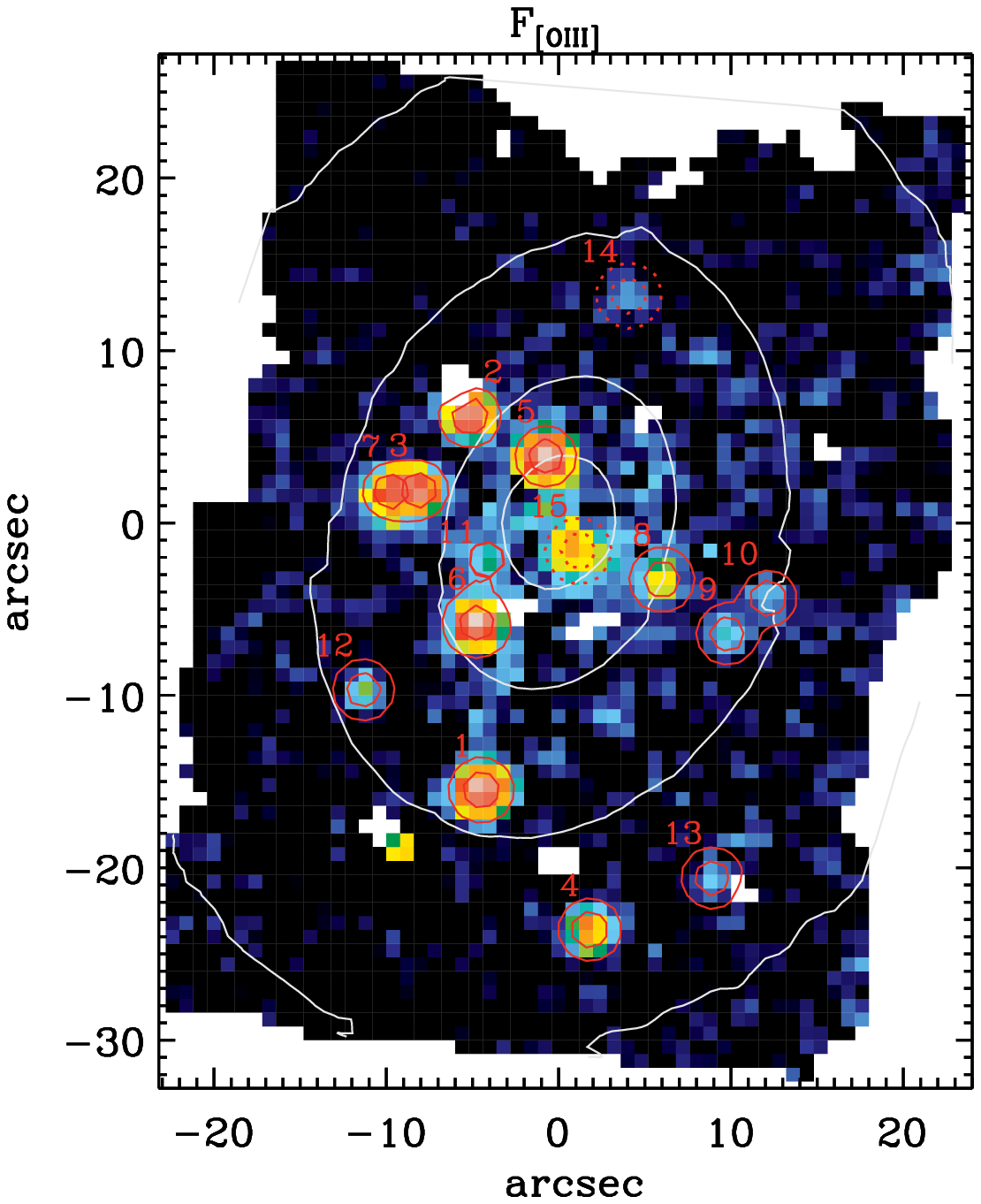}
\end{center}
\caption[]{Map of the flux of the \Oiii$\lambda5007$ emission from M32,
  in $10^{-16}\rm erg\,s^{-1}cm^{-2} arcsec^{-2}$ and in a logarithmic
  scale, together with the photometric measurements of both firmly and
  marginally detected PNe. Each of the PNe is labelled and delineated
  by red contours (in solid and dotted lines for solid and marginal
  detections, respectively) corresponding to the best-fitting Gaussian
  model to their observed flux distribution, which is used to compute
  their total \Oiii\ flux $F_{5007}$ and corresponding detection limit
  (see text). The inner contour around each PNe shows the half-peak
  flux level of the Gaussian model, thus corresponding to a circle a
  FWHM in diameter, whereas the second contour shows the region
  containing 90\%\ of the flux of the model around each PN, or of both
  the Gaussians models in the case of blended PNe sources.}
\label{fig:FluxPNe}
\end{figure}
}

\newcommand{\placefigFluxPNefits}{
\begin{figure*}
\begin{center}
\includegraphics[width=\textwidth]{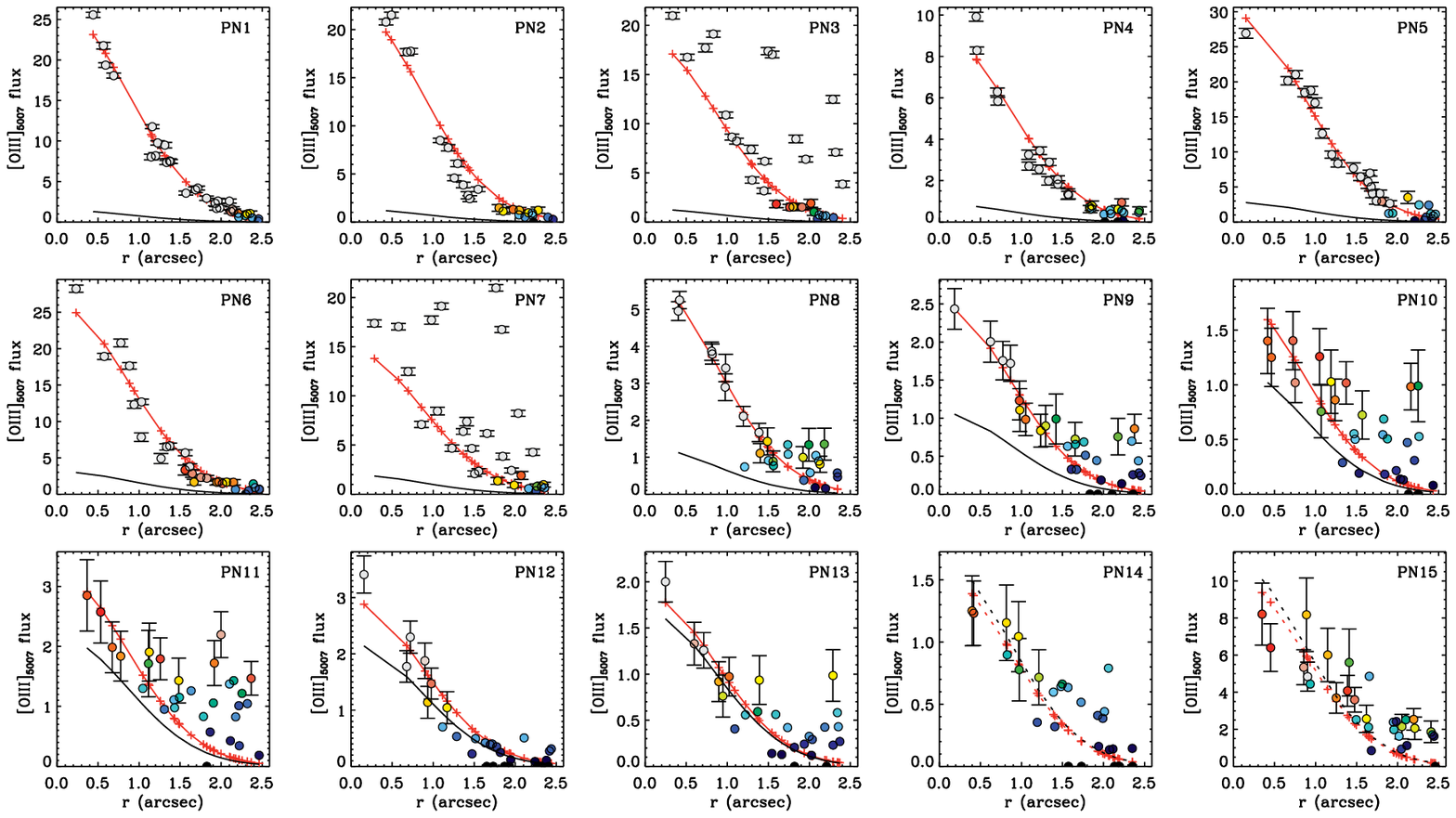}
\end{center}
\caption[]{Radial profiles of the flux of the \Oiii$\lambda5007$ line
  (in $10^{-16}\rm erg\,s^{-1}cm^{-2} arcsec^{-2}$) of the PNe sources
  shown in Fig.~\ref{fig:FluxPNe}, along with their corresponding
  best-fitting Gaussian model for the flux distribution (red lines,
  models are resampled in the \sauron\ $0\farcs8\times0\farcs8$
  bins). Radial distances are computed from the center of the Gaussian
  models and the data points are color coded according to the value of
  the $A/rN$ ratio. Colors change from blue to red for increasing
  values of $A/rN$ till they saturate to white for $A/rN > 6$. Green
  corresponds to the detection threshold of $A/rN=3$, above which the
  data points are also plotted together with error bars for the
  \Oiii\ flux. The black lines show the flux of the Gaussian model
  that would fit the data if these were rescaled until the minimum
  value of $A/rN$ within the FWHM (for a radius of 0\farcs96) of the
  model reaches the threshold of $A/rN=3$. When integrated, the
  best-fitting and rescaled Gaussian models yield the total flux
  $F_{5007}$ of each PN source and its corresponding detection limit,
  respectively. The model profiles for PNe 14 and 15 are shown with
  dashed lines to indicate that these are doubtful sources according
  the previous definition of the detection limit. The radial profiles
  extend out to a radius $r=3\sigma_{\rm PSF}$, where $\sigma_{\rm
    PSF}$ was determined from the fit of the brightest source PN
  1. PNe 3 and 7, 6 and 11, 9 and 10 have been modeled
  simultaneously.}
\label{fig:FluxPNefits}
\end{figure*}
}

\newcommand{\placefigFluxPNeSpectra}{
\begin{figure*}
\begin{center}
\includegraphics[width=\textwidth]{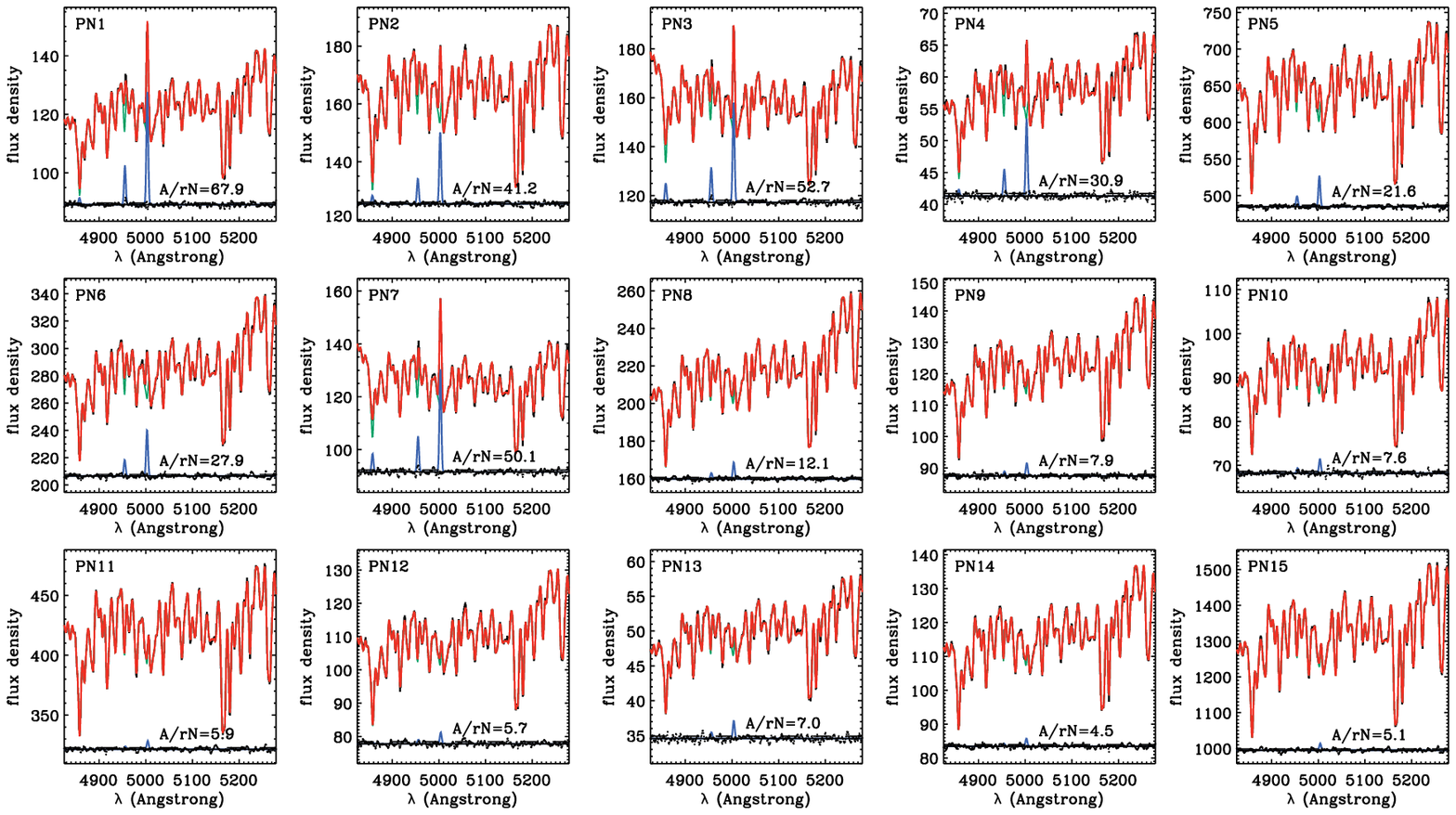}
\end{center}
\caption[]{\sauron\ aperture spectra for the PNe sources shown in
  Fig.~\ref{fig:FluxPNe} along with our best fit for the PNe emission
  and the background stellar continnum. Such spectra were extracted
  within a radius $r=3\,\sigma_{\rm PSF}$ from the center of the
  best-fitting Gaussian models plotted in Fig.~\ref{fig:FluxPNefits},
  and correspond to the sum of all the \sauron\ spectra observed in
  the bins shown in each of the panels of that figure. In each of the
  panels shown here the red line shows the sum of the stellar and
  nebular model (green and blue lines), with flux densities in
  $10^{-16}\rm erg\,s^{-1}cm^{-2} arcsec^{-2}\AA^{-1}$, whereas the
  small points show the fit residuals around a zero level that has
  been rescaled for clarity. The values of the $A/rN$ ratio for the
  \Oiii$\lambda5007$ line is printed at the bottom of each panel, and
  show that the \Oiii\ emission is detected in each spectrum,
  including those extracted around the doubtful PN sources 14 and 15.}
\label{fig:FluxPNeSpectra}
\end{figure*}
}

\newcommand{\placefigPNLF}{
\begin{figure}
\begin{center}
\includegraphics[width=\columnwidth]{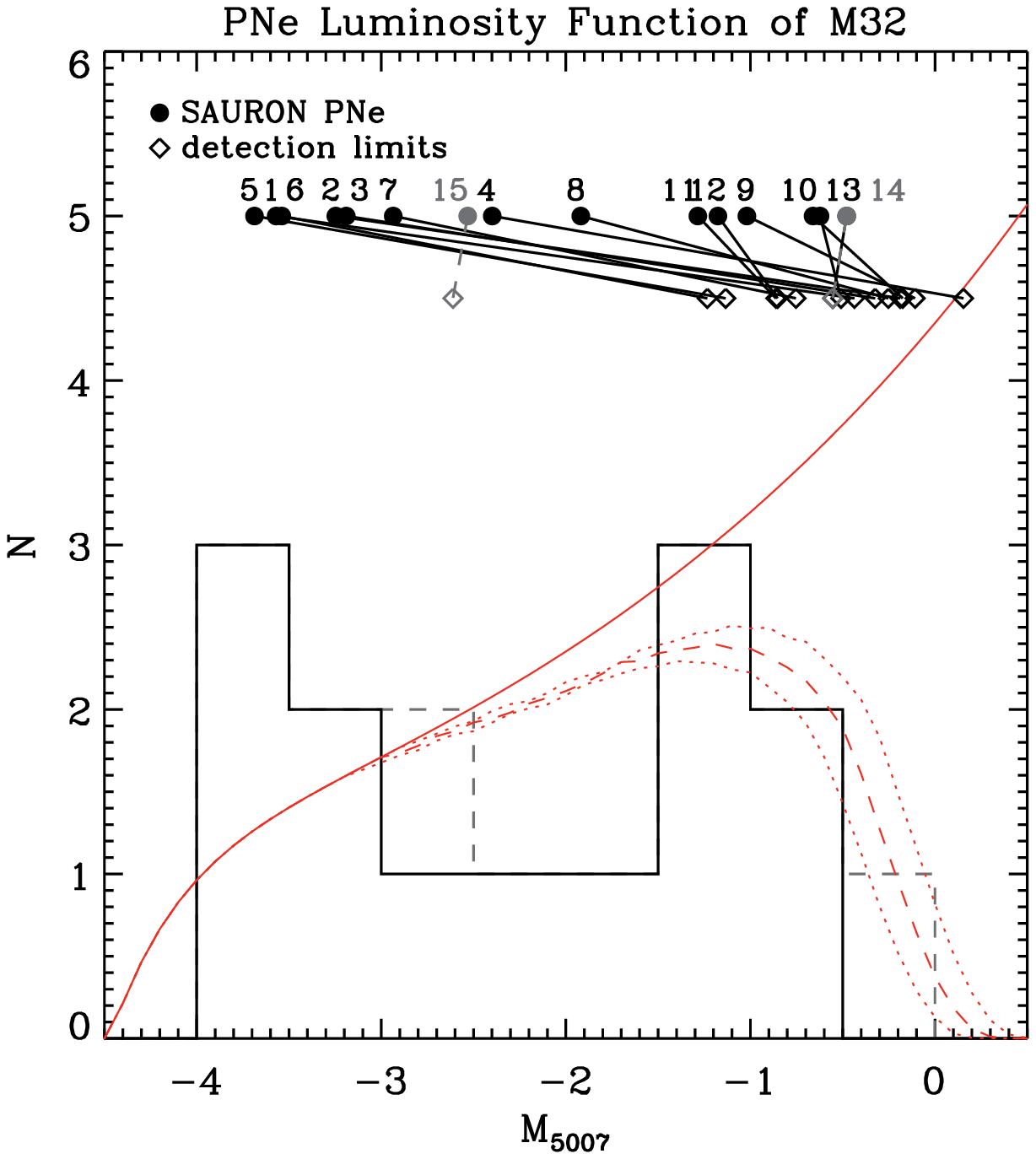}
\end{center}
\caption[]{Luminosity function of the PNe sources observed with
  \sauron\ in the optical regions of M32, including marginal
  detections (dashed gray histogram) and along with the theoretical
  form of the PNe luminosity function \citep[red solid line,
    from][]{Cia89}. The absolute magnitude of each PNe is shown at the
  top of the figure with filled circles that are plotted at an
  arbitrary constant ordinate and that are connected to open diamonds
  showing the magnitude corresponding to the detection limit of each
  source (with marginal sources shown in grey).
  The dashed line shows the PNe luminosity function multiplied by the
  median values of the completeness function across the entire
  \sauron\ field, whereas the dotted lines indicate the range by which
  the expected number of detected PNe would vary depending on their
  exact position within the $0\farcs8\times0\farcs8$ \sauron\ bins
  (see text and Fig.~\ref{fig:Completeness} for details).
  By integrating such a completeness-corrected luminosity function we
  obtain the total number of PNe that we would expect to detect within
  the region mapped by \sauron, and by matching this value with the
  actual number of observed PNe, 15 including marginal detections, we
  obtain our best estimate for the normalisation of the PNe luminosity
  function of M32.
  At first sight, the observed PNLF would look inconsistent with the
  theoretical prediction even accounting for incompleteness, in
  particular at its brightest end. A Kolgomorov-Smirnov test and the
  simulations shown in Fig.~\ref{fig:Simulations} demonstrate however
  that such a discrepancy can be expected with only 15 objects.}
\label{fig:PNLF}
\end{figure}
}

\newcommand{\placefigCompleteness}{
  \begin{figure*}
    \begin{center}
      \includegraphics[width=\textwidth]{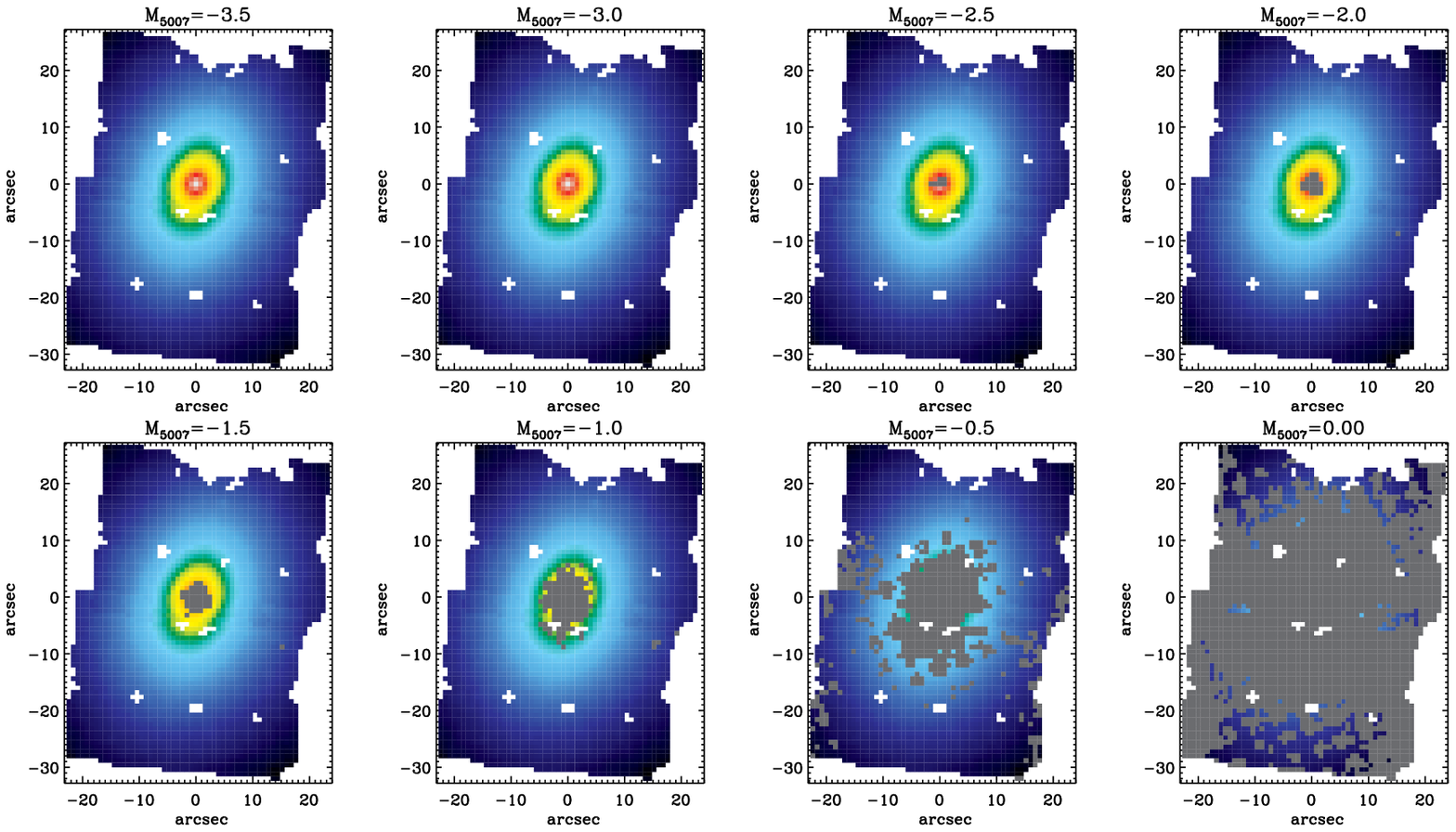}
    \end{center}
    \caption[]{Reconstructed optical images of M32, in logarithmic
      flux scale, showing with grey bins the regions where, from left
      to right, PNe of increasing absolute magnitude $M_{5007}$ and
      decreasing brightness would not be detected. To check whether in
      a given bin of the \sauron\ field a PN of absolute magnitude
      $M_{5007}$ would be detected, we generated at that position a
      Gaussian model for the \Oiii\ flux of a PN of that brightness in
      M32, deriving also the corresponding spectral density values for
      the amplitude of the \Oiii\ line. Using the values for the level
      of the noise $rN$ in the residuals of our spectral fits we then
      computed the values of the $A/rN$ ratio around the PN position
      and, using the criterion introduced in
      \S~\ref{subsec:AnalysisPNeDetection}, simply checked whether
      $A/rN > 3$ within a FWHM from the centre of the Gaussian model,
      which in these particular simulations correspond to the center
      of the \sauron\ bins.}
    \label{fig:Completeness}
  \end{figure*}
}

\newcommand{\placefigSimulations}{
  \begin{figure*}
    \begin{center}
      \includegraphics[width=\textwidth]{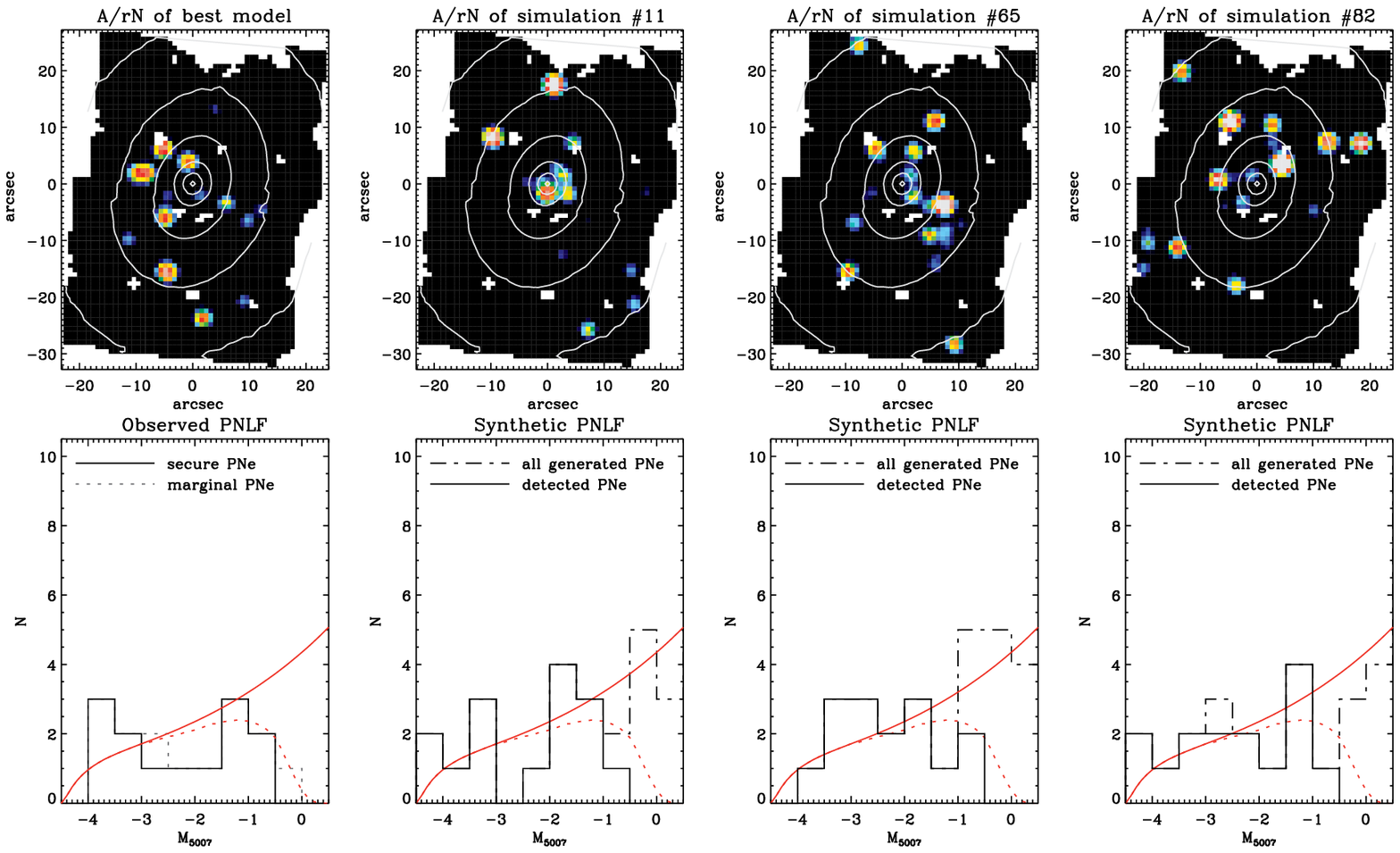}
    \end{center}
    \caption[]{Comparison between the spatial distribution and the
      luminosity function of the PNe detected in the central region of
      M32 (top and lower left panels, respectively) and three
      particular synthetic realisations of the same quantities
      (central and right panels).
      More specifically, the top panels show only the spatial
      distribution of PNe that were or would be detected in the
      optical regions of M32 by displaying values for the $A/rN$ ratio
      above the $A/rN> 3$ detection threshold, and which correspond
      (adopting the observed values of the residual-noise level $rN$)
      to the flux distribution of either our best-fitting Gaussian
      models to the PNe of M32 (\S\ref{subsec:AnalysisPNeDetection})
      or to similar Gaussian flux profiles for randomly generated PNe.
      Starting from an intrinsic \citeauthor{Cia89} shape of the PNLF
      (red curves in lower panels) and adopting a normalisation
      leading to match the number of observed PNe once the PNLF is
      corrected for incompletess (red dashed lines, see
      \S\ref{sec:Results}), the synthetic PN fields were generated by
      considering at any particular position in \sauron\ field of view
      the probability of having a PN of a given luminosity. Such a
      probability function simply corresponds to the total intrinsic
      PNLF rescaled by the fraction of stellar light that is observed
      in the \sauron\ spatial bin that is being considered.
      The \Oiii\ flux of the simulated PNe was then ``observed'' by
      obtaining maps for the $A/rN$ ratio and by applying the same
      detection criteria described in
      \S\ref{subsec:AnalysisPNeDetection}.
      These simulations illustrate that, given the low number of PNe
      expected in M32, not much can be read in the apparent bimodality
      of the PNLF that we observe nor in the fact that the PNe of M32
      seems to all be on the same side of the galaxy.
      In fact, these three particular simulations yielded the same
      total number of detected PNe than we observe (15), but generally
      this is not the case and the PNLF of such synthetic fields
      appear to differ even more from the observed PNLF than shown by
      the chosen examples.}
    \label{fig:Simulations}
  \end{figure*}
}

\newcommand{\placefigKinematics}{
\begin{figure}
\begin{center}
\includegraphics[width=\columnwidth]{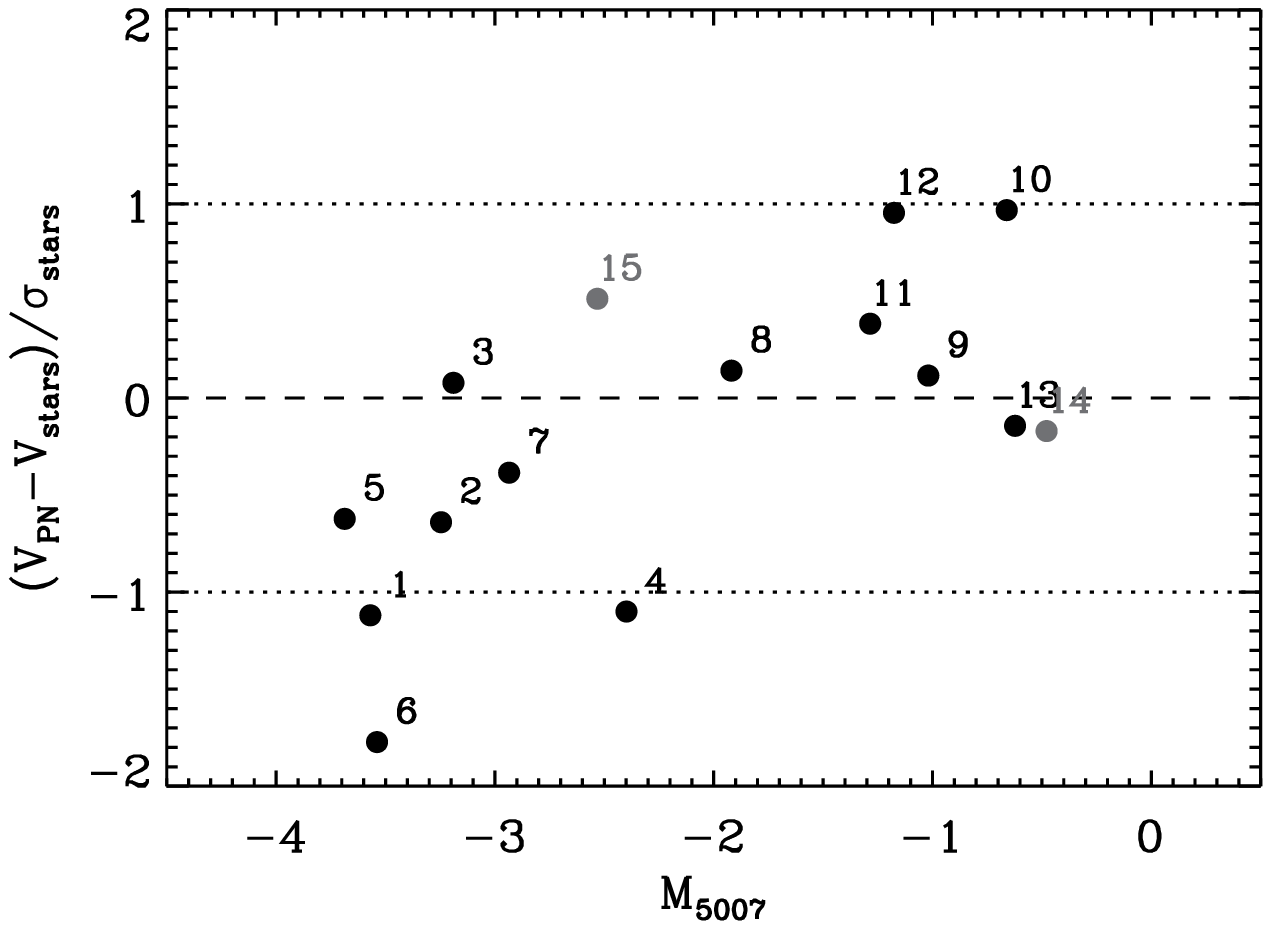}
\end{center}
\caption[]{Velocity $V_{\rm PN}$ of the PNe in M32 relative to the
  average stellar velocity $V_{\star}$ along the line-of-sight towards
  the PNe position and normalised by the stellar velocity dispersion
  $\sigma_{\star}$ in the same direction, sorted by their absolute
  magnitude $M_{5007}$. PNe sources are labelled and colour-coded as
  in Fig~.\ref{fig:PNLF}. The dotted horizontal lines for $|V_{\rm
    PN}-V_{\star}|/\sigma_{\star}=1$ delineate the region within which
  there is a 68\% probability that a given PN belongs to the stellar
  population of M32, assuming a Gaussian stellar line-of-sight
  velocity distribution. The $V_{\rm PN}$ values are plotted against
  $M_{5007}$ of the PNe to check whether bright and faint PNe,
  possibly powered by central stars having evolved via distinct binary
  and single-star channels, have different kinematics.
  A Spearman rank coefficient of 0.6 suggests a correlation in this
  figure, although this would not be dramatically significant since
  there is still a 2\% probability that the plotted quantities follow
  each other monotonically only by mere chance.}
\label{fig:Kinematics}
\end{figure}
}

\newcommand{\placefigUV}{
\begin{figure}
\begin{center}
\includegraphics[width=\columnwidth]{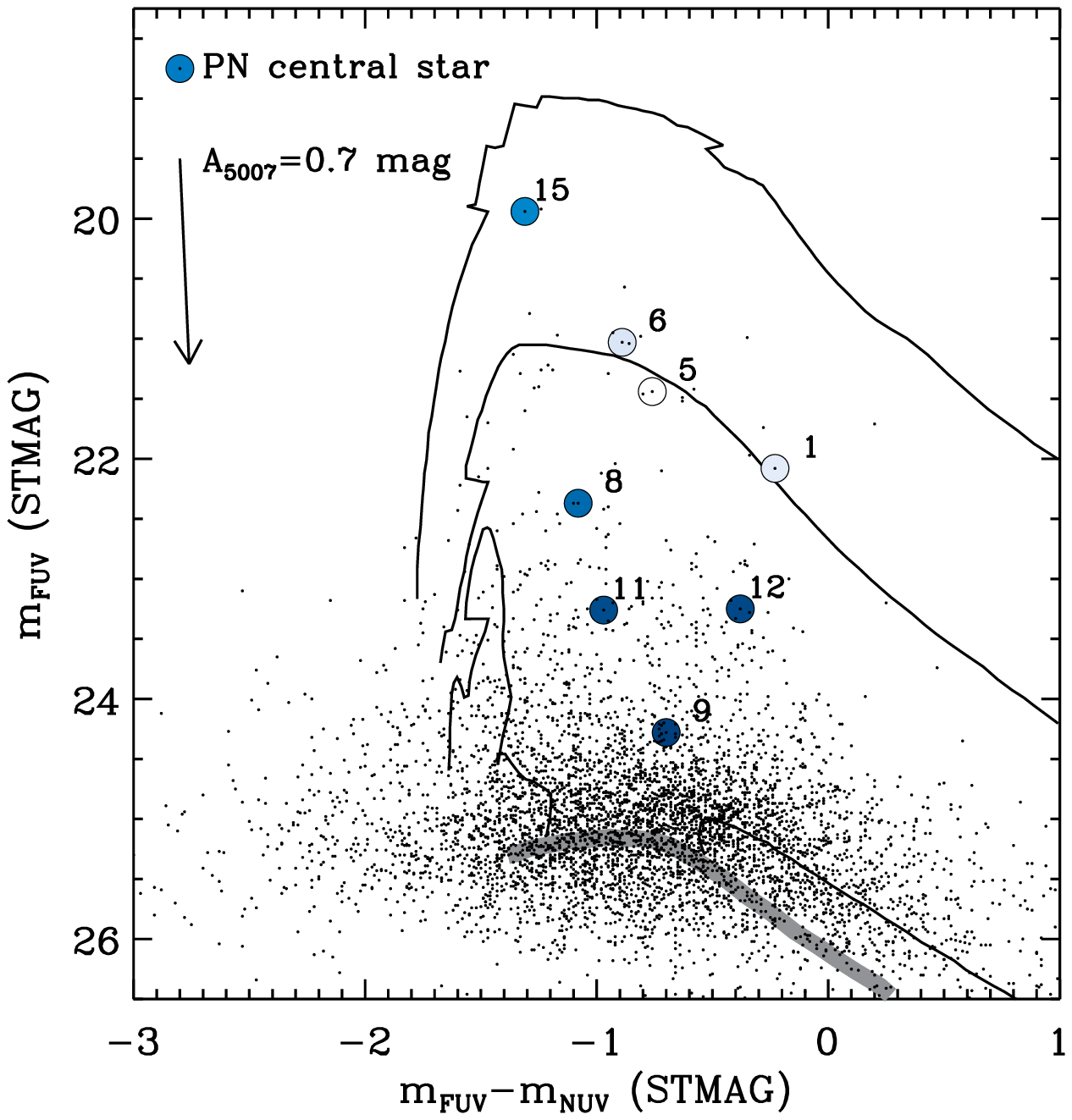}
\end{center}
\caption[]{ UV colour-magnitude diagram (CMD) for the HST data of
  \citet{Bro08} with solid lines tracing, from top to bottom, the
  evolutionary path of pAGB, early-pAGB, and AGB-manqu\'e stars
  starting from the horizontal branch (shown by the grey thick line)
  and ending in the region occupied by white dwarves. The asymptotic
  giant branch, from which pAGB and early-pAGB stars cross back the UV
  CMD is to the left and outside the plotting range.
  The large circles indicate the location in this CMD of the brightest
  and closest UV sources to the PNe falling within the field-of-view
  of the \citeauthor{Bro08} observations. Such PN central star
  candidates are colour-coded (from dark blue to white) according to
  the brightness of the PN they could be powering. Assuming a
  \citet{Car89} reddening law, the arrow at the top right of the
  figure shows the impact of dust reddening in the far- and near-UV
  HST passbands for an extinction of 0.7 magnitude at 5007\AA. This
  reddening value for PNe is typically observed in both star-forming
  and old galaxies, and would suffice to bring the candidate central
  stars of the PNe in the optical regions of M32 close to the location
  where pAGB and early-pAGB stars cross the UV CMD before reaching the
  WD cooling curve, to the left of the bluer end of the horizontal
  branch.}
\label{fig:UV}
\end{figure}
}

\newcommand{\placetabone}{
\begin{table}
\caption{Basic Properties of the PNe in the Optical Regions of M32}
\label{tab:PNeProp}
\begin{center}
\begin{tabular}{rrrrrrr}
\hline
ID  & x-off  & y-off  & $F_{5007}$ & $F_{\rm 5007, lim\/}$ & $V_{\rm PN}$ & \Oiii/\Hb \\ 
(1) & (2)    & (3)    & (4)        & (5)                   & (6)          & (7)       \\
\hline
1   &  -4.48 & -15.50 &      136.7 &                   7.6 & -214.2       &      11.5 \\
2   &  -5.16 &   6.18 &      101.5 &                   6.1 & -240.9       &      13.8 \\
3   &  -8.13 &   1.91 &       96.3 &                   6.9 & -190.2       &       6.3 \\
4   &   1.81 & -23.60 &       46.5 &                   4.4 & -216.1       &       7.8 \\
5   &  -0.74 &   3.86 &      152.3 &                  14.5 & -260.0       &      13.3 \\
6   &  -4.75 &  -5.83 &      132.9 &                  15.9 & -240.4       &      16.8 \\
7   &  -9.75 &   1.84 &       76.2 &                  10.2 & -225.6       &       7.7 \\
8   &   5.99 &  -3.29 &       29.9 &                   6.4 & -188.9       &        -- \\
9   &   9.78 &  -6.45 &       13.1 &                   5.6 & -188.1       &        -- \\
10  &  12.18 &  -4.38 &        9.4 &                   6.0 & -147.5       &        -- \\
11  &  -4.20 &  -2.09 &       16.7 &                  11.3 & -155.6       &        -- \\
12  & -11.33 &  -9.69 &       15.1 &                  11.2 & -133.6       &        -- \\
13  &   8.92 & -20.59 &        9.1 &                   8.2 & -189.1       &        -- \\
14  &   4.08 &  13.19 &        7.9 &                   8.5 & -231.6       &        -- \\
15  &   1.15 &  -1.62 &       52.6 &                  56.6 & -148.1       &        -- \\
\hline
\end{tabular}
\end{center}
Notes: 
(1)~PN ID.  
(2)--(3)~R.A. and Declination offset position, in arcseconds, from the center of M32.
(4)--(5)~Total \Oiii$\lambda5007$ flux and detection limit, in
$10^{-16}\rm erg\,s^{-1}cm^{-2}$.
(6)~Velocity in \kms, as measured in spectra extracted within a
FWHM wide aperture around the center of the best-fitting Gaussian
models shown in Fig.~\ref{fig:FluxPNe}.
(7)~\OiiioHb\ in the same spectra, when \Hb\ is detected.
As shown in Fig.~\ref{fig:Apertures} our sources 1 to 7 were already
detected by \citet{Cia89} who, in the order, labelled them with the ID
of 21, 26, 24, 20, 27, 23 and 25.
\end{table}
}

\title[PNe in the central regions of M32]{The Planetary Nebulae
  Population in the Central Regions of M32: the SAURON view}

\author[Sarzi et al.]{Marc Sarzi$^{1}$\thanks{E-mail :m.sarzi@herts.ac.uk}, 
Gary Mamon$^{2}$, Michele Cappellari$^{3}$, Eric Emsellem$^{4,5}$,
Roland Bacon$^{5}$, \newauthor Roger L.\ Davies, P.~Tim de Zeeuw$^{4,6}$\\
$^{1}$Centre for Astrophysics Research, University of Hertfordshire,
College Lane, Hatfield, Herts, AL10 9AB, UK\\
$^{2}$ Institut d'Astrophysique de Paris, 98 bis Bd. Arago, 75014
Paris, France\\
$^{3}$Sub-Dept of Astrophysics, Dept of Physics, University of Oxford,
Denys Wilkinson Building, Keble Road, Oxford, OX1 3RH, UK \\
$^{4}$European Southern Observatory, Karl-Schwarzschild-Str~2, 85748
Garching, Germany\\
$^{5}$Centre de Recherche Astronomique de Lyon, 9~Avenue Charles
Andr\'e, 69230 Saint Genis Laval, France\\
$^{6}$Sterrewacht Leiden, Universiteit Leiden, Postbus 9513, 2300 RA
Leiden, The Netherlands }

\begin{document}
\pagerange{\pageref{firstpage}--\pageref{lastpage}} \pubyear{2010}

\maketitle
\label{firstpage}

%
\begin{abstract}

Extragalactic Planetary Nebulae (PNe) are not only useful as distance
signposts or as tracers of the dark-matter content of their host
galaxies, but constitute also good indicators of the main properties
of their parent stellar populations.
Yet, so far, the properties of PNe in the optical regions of galaxies
where stellar population gradients can be more extreme have remained
largely unexplored, mainly because the detection of PNe with
narrow-band imaging or slit-less spectroscopy is considerably hampered
by the presence of a strong stellar background.
Integral-field spectroscopy (IFS) can overcome this limitation, and
here we present a study of the PN population in the nearby compact
elliptical M32.
Using \sauron\ data taken with just two 10-minutes-long pointings we
have doubled the number of known PNe within the effective radius of
M32, detecting PNe five times fainter than previously found in
narrow-band images that collected nearly the same number of photons.
We have carefully assessed the incompleteness limit of our survey, and
accounting for it across the entire range of luminosity values spanned
by our detected PNe, we could conclude despite having at our disposal
only 15 sources that the central PNe population of M32 is consistent
with the generally adopted shape for the PNe Luminosity Function and
its typical normalization observed in early-type galaxies.
Furthermore, owing to the proximity of M32 and to ultraviolet images
taken with the Hubble Space Telescope, we could identify the most
likely candidates for the central star of a subset of our detected PNe
and conclude that these stars are affected by substantial amounts of
circumstellar dust extinction, a finding that could reconcile the
intriguing discrepancy previously reported in M32 between the model
predictions and the observations for the later stages of stellar
evolution.
Considering the modest time investment on a 4m-class telescope that
delivered these results, this study illustrates the potential of
future IFS investigations for the central PNe population of early-type
galaxies, either with existing \sauron\ data for many more, albeit
more distant, objects, or from campaigns that will use the future
generations of integral-field spectrographs that will be mounted on
8m-class telescopes, such as MUSE on the Very Large Telescope.

\end{abstract}

%

%
\section{Introduction}
\label{sec:intro}

In the field of extra-galactic astronomy, Planetary Nebulae (PNe) are
perhaps regarded mostly either as useful indicators for the distance
of their galactic hosts \citep{Cia89,Jac90,Jac92}, thanks to the
almost universal -- though not fully understood -- shape of their
luminosity function (PNLF, generally in the \Oiii$\lambda5007$
emission), or as tracers of the gravitational potential in the
outskirts of galaxies \citep{Rom03,Dou07}.
Yet, as summarised in the review of \citet{Cia06}, PNe can also be
used as probes of their parent stellar population. For the closest or
brightest PNe, the detection of critical but weak diagnostic emission
lines such as \Oiii$\lambda4363$ allows to directly measure the
temperature of the ionised-gas and thus the metallicity of PNe, which
in turn makes it possible to constrain the chemical enrichment history
of their host galaxy \citep[e.g.,][]{Ric99,Jac99,Dop97}. In more
distant galaxies, it is still possible to place useful constraints on
the Oxygen abundance of PNe by using brighter lines such a
\Oiii$\lambda5007$, \Hb, \Nii$\lambda\lambda6548,6584$, \Ha\ or
\Sii$\lambda\lambda6713,6731$, or to investigate the star-formation
history of a galaxy by studying the shape and normalisation of the
PNLF \citep[e.g.,][]{Mar04,Sch07,Men08}. In fact, understanding the
origin of the PNLF is a puzzle that, once solved, promises to reveal
new clues on the late stages of stellar evolution and on the formation
of PNe themselves \citep[e.g.,][]{Cia05,Buz06}.

In this context it is important to note that most known extra-galactic
PNe have been found in the outskirts of their hosts, whereas the PNe
population of the optical regions of galaxies remains largely
unexplored. Most PNe studies have indeed been carried out through
narrow-band imaging \citep[e.g.,][]{Cia89} or slit-less spectroscopy
\citep[e.g.,][]{Dou07}, where the detection of PNe is considerably
hampered by the presence of a stellar background. Yet, it is in
central regions of galaxies where stellar population gradients can be
more extreme and where the stellar ages, metallicities and element
abundances can be the most diverse between different galaxies.
Integral-field spectroscopy can overcome the previous instrumental
limitations since it allows for the careful modelling of a galaxy's
integrated stellar spectrum, and this paper aims to demonstrate the
potential of IFS for studying the properties on PNe in the optical
regions of galaxies using \sauron\ observations for the compact
elliptical M32 (NGC221), our closest early-type galaxy.

This work is organized as follows. In \S\ref{sec:Data}, we briefly
review the acquisition and reduction of the \sauron\ data for M32. In
\S\ref{sec:Analysis} we detail how we optimised the extraction of the
nebular emission in the \sauron\ data and how we proceeded to identify
and measure the \Oiii$\lambda5007$ flux of the PNe in M32. In
\S\ref{sec:Results} we present our main results, assessing in
particular whether our data are consistent with the generally adopted
shape for PNLF and the value for its normalisation that is most
commonly observed in early-type galaxies. In \S\ref{sec:Discussion} we
discuss these findings by linking them to the known properties of the
stellar population of M32, and in particular to the apparent dearth of
post-asymptotic giant branch stars that was reported by
\citet{Bro00}. Finally, in \S\ref{sec:Conclusions} we draw our
conclusions and consider some future prospects for more IFS studies of
the PNe populations in the central regions of galaxies.

%
\section{Observations and Data Reductions}
\label{sec:Data}

M32 (NGC221) was one of the special objects that were observed over
the course of the \sauron\ representative survey \citep{deZ02}, and
more specifically during the last run of that observing campaign at
the 4m William Herschel Telescope and following the installation of a
new volume phase-holographic grating \citep{Ems04}. The central
regions of M32 were observed with two offset pointings during 600s for
each exposure, while using the low-resolution mode of \sauron\ that
gives a field of view of $33\farcs0\times44\farcs0$ fully sampled by
$0\farcs94\times0\farcs94$ square lenses \citep[for more details on
  the instrument see][]{Bac01}. The data from each pointing were
reduced similarly to the data obtained for the objects of the main
\sauron\ sample \citep[see][]{Ems04,Fal06}, and the resulting
datacubes were merged and resampled in $0\farcs8\times0\farcs8$
spatial elements each corresponding to spectra with a final spectral
resolution of 4.2\AA\ (FWHM).

The present \sauron\ data for M32 were already used to extract the
stellar kinematics that was modeled by \citet{Cap06} and the only
difference with the data used in that work and other papers of the
\sauron\ project is that here we did not perform any Voronoi spatial
binning \citep{Cap03}. This was done to avoid swamping the signal of
the weaker PNe against an increased stellar background, and to allow
for a more consistent analysis of the flux distribution from each of
the unresolved PN that we may detect.

\placefigApertures

%
\section{Data Analysis}
\label{sec:Analysis}

\subsection{Emission-line Measurements}
\label{subsec:AnalysisEmission}

In order to identify PNe in M32 and measure their flux in the
\Oiii$\lambda5007$ line we first need to separate as accurately as
possible the stellar and nebular contribution to each of the
\sauron\ spectra that sample the central regions of this galaxy.
For this purpose we used the method of \citet[][hereafter Paper~V
  following the notation of the \sauron\ project]{Sar06} whereby a set
of stellar templates and Gaussian emission lines are fitted
simultaneously to the spectra\footnote{
In practice this is achieved by using the IDL code GANDALF (available
at {\tt http://star-www.herts.ac.uk/$_{\widetilde{~}}$sarzi\/}) and
the stellar kinematics extracted with the pixel-fitting IDL code pPXF
\citep[][{\tt
    http://www-astro.ox.ac.uk/$_{\widetilde{~}}$mxc/idl\/}]{Cap04}
}, while following also the approach of \citet{Sar10} to further
improve the match to the stellar continuum and ensure that the
ionised-gas emission is extracted from the subtraction of a physically
motivated stellar model.
This is achieved by using, instead of standard template libraries
based on stellar spectra or single-age stellar population models, a
more appropriate set of empirical templates that are constructed while
matching a number of high-quality \sauron\ spectra extracted from
regions in the target galaxy where no ionised-gas emission is found.

\placefigAoN
\placefigFluxPNe
\placefigFluxPNefits
\placefigFluxPNeSpectra

Fig.~\ref{fig:Apertures} shows how we identified such emission-line
free regions in M32 using an \Oiii\ narrow-band image made from the
\sauron\ data themselves (see the caption of Fig.~\ref{fig:Apertures}
for details). Each of the spectra extracted from the circular
apertures shown in Fig. 1 was fitted with the pixel-fitting code of
\citet{Cap04}, over the full wavelength range of the \sauron\ data and
using the entire MILES template library of \citet{San06}.
Both the quality of such aperture spectra and of our fit to them is
very high, with values for the ratio between the median level of the
spectra ($S$) and the average level of the fit residuals (residual
noise, $rN$) ranging from 100 to over 400, with $S/rN\sim300$ on
average.
The weights assigned to the MILES stellar templates during each of
these fits were then used to combine the MILES spectra into optimal
templates that, owing to the excellent quality of our fit, can be in
practice regarded as M32 spectra deconvolved from the line-of-sight
kinematical broadening.
Fig.~\ref{fig:Apertures} also illustrates how our emission-free
apertures are evenly spread over the \sauron\ field of view, which
ensures that their corresponding optimal templates can account for the
presence of stellar population gradients \citep{Ros05} when they are
used to match each of the single spectra sampling the central regions
of M32.

Fig.~\ref{fig:Apertures} also locates the PNe found in the narrow-band
survey of \citet{Cia89}, and reveals already the presence of weaker
and isolated peaks of 5007\AA\ flux that may also originate from the
unresolved \Oiii\ emission of PNe.
If this is the case, then the nebular spectra observed where the flux
of an unresolved PN is scattered by the atmospheric point-spread
function (PSF) should be characterised by large values of the
\OiiioHb\ ratio, virtually no \Ni$\lambda\lambda5197,5200$ emission,
and by very narrow lines that would not be resolved in the
\sauron\ spectra.
The standard fitting strategy of Paper~V, whereby the \Oiii\ doublet
is fitted first and the \Hb\ flux is subsequently extracted holding
the \Hb\ line profile to that of the \Oiii\ lines, is therefore
appropriate for isolating the emission of PNe, and here we only
further imposed a fixed width on the \Oiii\ lines, for an observed
velocity dispersion of 108 \kms\ that corresponds to the spectral
resolution of the \sauron\ data (\S~2).

Fig.~\ref{fig:AoN} shows the map of the ratio between the best-fitting
Gaussian amplitude $A$ for the \Oiii$\lambda5007$ line and the noise
level $rN$ in the fit residuals.  In Paper~V we concluded that a
minimum value of $A/rN=4$ was required to detect \Oiii\ emission,
although this was while letting the width of the lines free to vary.
Since instead here we have fixed the \Oiii\ line width, our fits
involve one less free parameter and we can measure unbiased
\Oiii\ fluxes also down to a $A/rN=3$ threshold, similarly to the case
of \Hb\ line measurements of Paper~V when the \Oiii\ line profile was
imposed on the \Hb\ line.
With such a detection limit, Fig.~\ref{fig:AoN} confirms the presence
of several isolated weak patches of detected \Oiii\ emission beside
the PNe of \citet{Cia89}, and shows the absence of a diffuse
ionised-gas component that is otherwise commonly found in early-type
galaxies (Paper~V).

\subsection{PNe Detection and Flux Measurements}
\label{subsec:AnalysisPNeDetection}

To measure the flux of any known or possible PNe in the
\sauron\ field-of-view, and at the same time establish whether a weak
patch of \Oiii\ emission is indeed consistent with the flux
distribution expected from an unresolved PN, we fitted each of the
\Oiii\ sources in Fig.~\ref{fig:AoN} with a circular Gaussian function
that is meant to represent the shape of the PSF of our observations.
The full-width at half maximum (FWHM) of such a PSF was first measured
by matching the most cleanly detected of our sources
\citep[corresponding to source 21 of][]{Cia89}, and subsequently held
constant when measuring the flux of other known or candidate PNe. The
Gaussian model was resampled in \sauron\ $0\farcs8\times0\farcs8$ bins
before comparing it to the data, and for close PN sources care was
taken to deblend their fluxes by fitting simultaneously two or more
Gaussian functions.
Based on these fits, we deemed a given source of \Oiii\ emission a
detected PN if $A/rN>3$ within at least the FWHM of the Gaussian
model. The model itself was then used to measure the \Oiii\ flux of
the PN, by integrating the entire Gaussian function out to radii where
otherwise the \Oiii\ emission generally would not be detected or could
come from other nearby PNe.

According to the previous detection criterion we find 13 PNe in the
central regions of M32 observed by \sauron, with two additional
sources that barely miss the detection threshold due to just one value
of $A/rN$ being less then 3 within the FWHM of the models, which
turned out to be 1\farcs94 for a $\sigma_{\rm PSF\/}=0\farcs82$.
Each of these sources is identified and labelled in
Fig.~\ref{fig:FluxPNe}, which maps the \Oiii\ flux observed in M32 and
shows with red contours the best fitting Gaussian models to each of
the detected PNe, including the questionable sources with dotted
lines.
The fit to the \Oiii\ flux distribution of these sources can be better
appreciated in Fig.~\ref{fig:FluxPNefits}, where the radial profile
for the \Oiii\ flux of each PN is plotted together with its
best-fitting Gaussian model.
In Fig.~\ref{fig:FluxPNefits} the \Oiii\ data points are colour-coded
according to the value of the $A/rN$ ratio measured in each bin to
illustrate that down to our \Oiii\ detection threshold of $A/rN=3$ the
\Oiii\ fluxes follow very well the PSF model, which provides an
independent check of our $A/rN$ detection criterion for the
\Oiii\ emission.

Fig.~\ref{fig:FluxPNefits} also shows how for each source we determine
the minimum flux that we could have measured at the particular
position of the PN. In fact, we can expect such detection limits to
vary across the galaxy, and more specifically to increase towards the
center where only the brightest PNe lead to \Oiii\ lines of sufficient
strength to stick out above the higher level of statistical noise that
comes with a brighter stellar continuum.
Following our detection criterion for PNe, we set the detection limit
at the location of each our PN sources as the flux of the Gaussian
model (shown with a black line in Fig.~\ref{fig:FluxPNefits}) that
would match the \Oiii\ fluxes if these, and the corresponding values
of the \Oiii$\lambda5007$ line amplitude $A$, would be rescaled until
the minimum $A/rN$ value inside a FWHM reaches the threshold of
$A/rN=3$.
As expected, for the doubtful sources the detection limits exceed the
observed flux of the best model, since the minimum $A/rN$ value is
already less than 3 and therefore the data have to be scaled up,
instead of down (although by just $\sim$7\% in both cases).
The significance of the detection of our PNe sources determines the
order in which the PNe are labelled in Figs.~\ref{fig:FluxPNe},
\ref{fig:FluxPNefits} and subsequent figures.

To further illustrate the degree to which all the PNe shown in
Fig.~\ref{fig:FluxPNe} are detected, for each source we added all the
\sauron\ spectra within a radius $r=3\,\sigma_{\rm PSF}$ from the
center of the best-fitting Gaussian models (corresponding to all the
bins plotted in the panels of Fig.~\ref{fig:FluxPNefits}), and fitted
the resulting spectra with our standard procedure.
Fig.~\ref{fig:FluxPNeSpectra} shows such aperture spectra and our best
fit to both the nebular emission from each PN and to the stellar
continuum observed along the line-of-sight towards them.
We note that the $A/rN$ values for the \Oiii\ emission measured in
these spectra always indicate a detection, even within a
$3\,\sigma_{\rm PSF}$ aperture that is 6.5 times wider than the FWHM
region where we decided to assess the detection of PNe and within
which we might have expected the emission from the weakest of our PNe
to be lost against a larger stellar background.
That this is never the case, however, suggests that the two
questionable sources might as well be regarded as marginal detections,
which is why we will keep considering them in the remainder of the
paper.

To conclude this section in Tab.~\ref{tab:PNeProp} we list, for both
firmly and barely detected PNe in M32, the position relative to the
center of the galaxy, the total $F_{5007}$ flux of the \Oiii\ emission
with its corresponding detection limit, the velocity of each source,
and, when the \Hb\ line was also detected, the average value of the
\OiiioHb\ ratio.
The latter two measurements are based on fits to spectra similar to
those presented in Fig.~\ref{fig:FluxPNeSpectra}, but now extracted
within a FWHM-wide aperture (rather than within a radius
$r=3\,\sigma_{\rm PSF}$) in order to maximise the emission-line signal
and better isolate the kinematics of PNe that are close to each other
in projection.

\placetabone 

%
\section{Results}
\label{sec:Results}

The analysis described in the previous section has delivered the solid
detection of 13 PNe within the optical regions of M32, with an
additional 2 sources where the observed \Oiii\ flux distribution is
only marginally consistent with the emission from an unresolved PN.
Given the systemic velocity $V_{sys}=-197\,\kms$ of M32 and its
average stellar velocity dispersion within one effective radius
$\sigma_e = 60\,\kms$ \citep{Cap06}, the velocities listed in
Tab.~\ref{tab:PNeProp} show that the observed PNe very likely belong
to M32, rather than to the disk or halo of M31 where PNe move on
average at a speed of $-400\,\kms$ \citep{Mer06}.
In fact, the case for membership in M32 holds even at a {\it local\/}
level, when the PNe velocity $V_{\rm PN}$ is compared with the mean
stellar velocity $V_{\star}$ and velocity dispersion $\sigma_{\star}$
measured along the line-of-sight towards their location.
Using FWHM-wide aperture spectra to better separate the velocity
$V_{\rm PN}$ of blended or close PNe and larger $3\,\sigma_{\rm
  PSF}$-wide apertures to extract robust $V_{\star}$ and
$\sigma_{\star}$ measurements also in the outskirts of M32, we found
that only in 3 cases $| V_{\rm PN} - V_{\star} | > \sigma_{\star}$,
even less often than what is normally expected for 15 sources.
Tab.~\ref{tab:PNeProp} also lists, when we could measure it, large
values for the \Oiii/\Hb\ ratio, thus showing that these are indeed
high-ionisation sources most likely consistent with PN spectra.
Finally, we note that for the PNe that were also detected by
\citet{Cia89} our flux values agree fairly well, within a few percent,
with the values reported in that work. For instance, for the brightest
PN in our sample (source 5) we measure a $F_{5007}$ flux that is only
8\% fainter than what was found by \citet[][source 27 in their
  Tab.~7]{Cia89}.

\placefigPNLF
\placefigCompleteness

Fig.~\ref{fig:PNLF} presents the luminosity function of the PNe found
in the optical regions of M32 by our \sauron\ observations, which
essentially map this galaxy within its effective radius $R_e =
30\arcsec$.
To construct such a PNLF we followed the definition of \citet{Cia89}
to compute the apparent V-band magnitude $m_{5007} =
-2.5\log{F_{5007}} -13.74$ of our sources, and derived the values for
the absolute magnitude $M_{5007}$ that are shown in
Fig.~\ref{fig:PNLF} by adopting a distance modulus of 24.49
magnitudes.
This corresponds to a distance of 791 kpc, and it is the same value
based on the surface-brightness fluctuation measurements of
\citet{Ton01} that was used by \citet{Cap06}.

Fig.~\ref{fig:PNLF} also shows with a red line the expected form for
the PNLF introduced by \citeauthor{Cia89}, who combined the simple
\citet{Hen63} model for a population of slowly fading and expanding
PN envelopes with a sharp exponential cutoff at the high-mass end.
The shape of our observed PNLF is inconsistent with such a theoretical
curve, but this is hardly surprising considering that at the faint end
our PN number counts are affected by incompleteness.
To a first approximation the onset of this bias can be appreciated by
noticing in Fig.~\ref{fig:PNLF} how the values for the absolute
magnitude corresponding to the detection limits of each of our sources
(shown with open diamonds) pile up from a $M_{5007}\sim-1.5$ and
brighter , corresponding to an apparent magnitude limit of
$m_{5007}\sim23$.
In general, while looking at bigger galaxies than M32 with a
correspondingly larger number of PNe or at very close objects such as
the large and small Magellanic clouds, other studies use the PNe
that are found within the completeness limit in order to determine
both the best normalisation for the theoretical PNLF and to test
whether this could in fact be considered the parent distribution for
the data.
In the case of M32, however, there are simply not be enough PNe
within the completeness limit (say for $M_{5007}<-2$, taking a
conservative guess from Fig.~\ref{fig:PNLF}) to confidently carry out
such a measurement and test.
We therefore decided to first understand how our observational biases
would affect the theoretical form of \citeauthor{Cia89} for the PNLF,
to then test whether {\it all \/} our data, over the entire $M_{5007}$
range, could have been drawn from the corresponding
completeness-corrected model PNLF, while also determining the best
normalisation for it.

Fig.~\ref{fig:Completeness} presents simulations designed to show how
we derive the completeness function of our observations, that is, the
probability as a function of absolute magnitude $M_{5007}$ that a PN
of that brightness could be detected across the {\it entire\/}
field-of-view of the \sauron\ observations for M32.
Since at any given luminosity the number of expected PNe scales with
the total number of stars, such a probability is just the fraction of
the total stellar flux encompassed by our field-of-view that is
observed within the area over which PNe of a given absolute magnitude
$M_{5007}$ could be detected.
To then decide whether a PN of given $M_{5007}$ could be detected at
any given position, we generated a Gaussian PN model of total flux
corresponding to the apparent magnitude $m_{5007}$ and, using the
values of $rN$ obtained from our spectral fits across the entire
field-of-view, simply checked whether $A/rN > 3$ within a FWHM of the
Gaussian model.
As Fig.~\ref{fig:Completeness} illustrates, starting from an absolute
magnitude of $M_{5007}\sim-2.5$ ($m_{5007}\sim22$) it becomes more and
more difficult to detect PNe near the central region of the galaxy,
with the area of non-detectability (shown with grey bins) quickly
expanding outward for PNe fainter than $M_{5007}\sim-1.0$
($m_{5007}\sim23.5$).
More specifically, we are 100\% complete down to $M_{5007}\sim-3$,
whereas we lose about 5\% of PNe with an absolute magnitude of
$M_{5007}\sim-2.5$ and nearly 26\% of those with $M_{5007}\sim-1.0$.
Fig.~\ref{fig:Completeness} shows simulations for just eight
$M_{5007}$ values and while placing the Gaussian PN models at the
center of each \sauron\ $0\farcs8\times0\farcs8$ bin, but to derive
the completeness function and the uncertainties associated with it we
adopted a more refined grid of $M_{5007}$ values and accounted for the
impact of randomly placing the Gaussian models within the
\sauron\ bins.

When the theoretical PNLF of \citeauthor{Cia89} is multiplied by the
completeness function that we have just derived, we finally obtain the
completeness-corrected prediction for the expected number of PNe in
M32 that is shown by the red dashed curve in Fig.~\ref{fig:PNLF}, with
the dotted red lines tracing the scatter in this correction due to the
exact PNe position in the \sauron\ bins.
A Kolgomorov-Smirnov test reveals that this corrected PNLF can be
regarded as the parent distribution for our data, since there is a
82\% probability that all of our 15 PNe could have been drawn from it,
and a 52\% probability when considering only the 13 secure PNe.
By integrating the completeness-corrected luminosity function we
obtain the total number of PNe that we would expect to detect, and by
matching this value with the actual number of observed PNe, 15
including marginal detections, we obtain our best estimate for the
normalisation of the PNe luminosity function of M32 within the region
mapped by our \sauron\ observations.
Such a normalisation is generally expressed in terms of the
luminosity-specific number density of PNe, through the ratio
$\alpha=N_{\rm PNe}/L_{\rm bol}$ between the total expected number of
PNe $N_{\rm PNe}$ in a stellar population and the total bolometric
luminosity $L_{\rm bol}$ of the latter.
Typically, the total number of PNe is estimated by integrating the
normalised PNLF of \citeauthor{Cia89} from the brightest observed
absolute magnitude $M_{\star,5007}=-4.47$ to 8 magnitudes fainter,
which is the limit where \citet{Hen63} locate the faintest PNe.
Yet, since the faint end of the PNLF has not been well
constrained by observations, in some instances $\alpha$ is provided
only within the completeness limit and such an extrapolation is
avoided.
Most often in the literature this limit lies within 2.5 magnitudes of
$M_{\star,5007}$ and the luminosity-specific number density parameter
is therefore indicated as $\alpha_{2.5}$.
Adopting our normalisation there should be 76 PNe in total within the
central region of M32 mapped by \sauron\ and between 7 and 8 PNe (7.6)
with $M_{5007} - M_{\star,5007} \le 2.5$. 
Given that our observations cover M32 within essentially its effective
radius $R_e$ from I-band images, we are here encompassing nearly half
of the total I-band luminosity, i.e. $L_I=2.3\times10^8M_{\odot}$
\citep[see][for $R_e$ and total luminosity I-band
  measurements]{Cap06}.
Adopting the reddening corrected $B-V=0.88$ colour of \citet{Buz06}
for M32, and using \citet{Bru03} models for a Salpeter IMF and Solar
metallicity, we then obtain a $B-I=2.01$ and thus a B-band luminosity
of $L_B=1.3\times10^8M_{\odot}$.
Finally, following \citet{Buz06} also for the bolometric correction to
the B-band luminosity we arrive at a bolometric luminosity $L_{\rm
  bol}=3.4\times10^8M_{\odot}$, which means that for the optical
regions of M32, within $R_e$, the luminosity-specific PN number
density is $\alpha=2.2\times10^{-7}{\rm L}_{\odot}^{-1}$, and a factor
ten less for $\alpha_{2.5}$.
Considering a 33\% statistical error associated with our normalisation
based on 15 points (for an intrinsic number of 25 PNe within our
detection magnitude range), these values are remarkably consistent
with thefindings of both \citet{Buz06} and \citet{Cia89} who give
$\alpha=1.70\times10^{-7} {\rm L}_{\odot}^{-1}$ and
$\alpha_{2.5}=2.23\times10^{-8} {\rm L}_{\odot}^{-1}$, respectively.

To conclude, we have detected 15 PNe in the optical regions of M32
that were mapped by our \sauron\ observations, and thanks to the
integral-field nature of our data and by carefully accounting for the
incompleteness of our observation we have been able not only to
confirm kinematically that these sources belong to M32 but also that
their observed luminosity function and their total number are
consistent with the generally adopted shape of the PNLF \citep[that
  of][]{Cia89} and the typical values for the PNe number density
$\alpha$ in early-type galaxies \citep[a few $10^{-7}$ per ${\rm
    L_{\odot}}$,][]{Buz06}.
Considering that our 600s-long \sauron\ observations at the 4m William
Herchel Telescope collected nearly as many photons as the 3600s- and
1800s-long narrow-band observations of \citet{Cia89} at the 0.9m and
2.1m Kitt Peak telescopes, respectively, the fact that we more than
doubled the number of known PNe in the central regions of M32 while
reaching 5 times fainter \Oiii\ PNe fluxes (see Tab.\ref{tab:PNeProp})
illustrates well the power of integral-field spectroscopy to study PNe
in the optical regions of galaxies.

\placefigSimulations

%
\section{Discussion}
\label{sec:Discussion}

It is interesting to comment on our PN results in light of what is
known about the central stellar population of M32 and the current
explanations for both the shape and normalisation of the PNLF.

\citet{Ros05} have shown that the luminosity-weighted mean stellar
population of M32 at 1$R_e$ is older by $\sim$3 Gyr and more
metal-poor by about $-0.25$ dex in [Fe/H] than the central value of
$\sim$4 Gyr and [Fe/H]$\sim0.0$, which are trends that conspire to
produce the flat color profiles previously reported in this object
\citep{Pel93,Lau98}.
Episodes of star formation may enhance for a relative short period of
time the bright-end of the PNe luminosity function of a galaxy, since
massive main-sequence stars tend to produce also massive and bright
central PNe stars \citep[see e.g., the case of M33 in][]{Cia04}.
For instance, \citet{Mar04} shows that stars with initial mass between
2.5 and 3.5$\rm M_{\odot}$ evolve to central PN stars of about $0.70 -
0.75\rm M_{\odot}$ that can power nebular fluxes corresponding to the
typical PNLF cut-off magnitude $M_{\star,5007}$ of -4.47.
Since massive stars are short-lived, however, such a PNLF enhancement
would be pretty weak already 3 Gyr after star formation ceased, and it
would no longer apply to the brightest PNe but to objects with
$M_{5007}\sim -2.5$ \citep{Mar04}. It is thus not surprising, in
particular given the small number of PNe at our disposal, that we do
not see such a signature in our observed PNLF. If anything, there
seems to be a lack of PNe around that intermediate $M_{5007}$ value.

On the other hand, and like in other early-type galaxies with no
evidence of recent (less than 1-Gyr-old) star formation, the presence
in M32 of bright PNe at a time when the progeny of massive stars have
long disappeared is a puzzle that has not yet been fully solved.
In fact, the need for a massive central star in the brightest PNe,
with mass above $\sim0.70 M_{\odot}$, is a result that still holds in
the studies of \citet{Sch07} and \citet{Men08} that supersede the
initial investigation of \citet{Mar04}.
One possible way to form the required high-mass PN cores, suggested by
\citet{Cia05}, involves the binary coalescence during their hydrogen
burning phase of $\sim1\rm M_{\odot}$ stars, a process that has been
associated with the formation of blue-stragglers and which could be
facilitated in early-type galaxies by the abundance of $\sim1\rm
M_{\odot}$ stars in old stellar systems.
If such a second path to the formation of PNe exists, then we may also
expect to find a PNLF characterised by a peak at high $M_{5007}$ above
an otherwise monothonically increasing PNLF, corresponding to a
standard \citeauthor{Hen63} population of PNe powered by only old
central PN stars of relatively low mass.
In light of this scenario, it would be tempting to see a bimodality in
our observed PNLF (Fig.~\ref{fig:PNLF}), which happens to show two
peaks around $M_{5007}\sim$-3.5 and -1 that correspond to parent
stellar masses near 2 and 1$\rm M_{\odot}$, respectively.
However, due to small number statistics we cannot read too much in the
observed shape of the M32 PNLF, as it is instructive to see through
simple simulations such as those presented in
Fig.~\ref{fig:Simulations}, which confirm the formal consistency
indicated by a simple KS-test between our data and the
completeness-corrected \citeauthor{Cia89} form of the PNLF.
Similarly, it would be difficult with our sample size to claim a
significant difference in the kinematic behaviour of bright and faint
PNe \citep[as in NGC~4697, ][]{Sam06}, which could occur for instance
if the coalesced central stars of the PNe can form in globular
clusters, as could be the case for X-ray binaries \citep{Sar00,Whi02}
and where indeed blue-stragglers are often found.
Alternatively, and consistent with the hypothesis of \citet{Mar04}, a
different kinematics for the brightest PNe may be a sign of recent
accretion of a younger and smaller galaxy \citep{Mam05}.
Yet, integral-field data allow for a direct comparison with the
stellar kinematics as a function of $M_{5007}$, such as shown for M32
in Fig.~\ref{fig:Kinematics}, that would be interesting to apply to
more massive systems hosting a larger number of PNe.

\placefigKinematics

From a theoretical perspective, our best normalisaton of the PNLF in
the optical regions of M32, with a value for the specific PN number
density $\alpha$ in line with that of other early-type galaxies, also
poses a problem.
Indeed, as long as the time-scale for PN visibility depends mostly on
the lifetime of their post-AGB cores (pAGB), then since cores of lower
mass spend more time in their high-temperature regime $\alpha$ should
increase with the stellar population age, whereas in fact quite the
opposite is observed \citep[see][for a comprehensive review of this
  discrepancy]{Buz06}.
A possible solution to this puzzle is to assume that a considerable
fraction of the stellar population of early-type galaxies ends up in
the extreme hot end of the helium-burning horizontal-branch (HB), and
that when such stars of very small hydrogen-envelope mass subsequently
leave the HB they not only skip entirely their AGB phase (thus the
name of AGB-manqu\'e) but they also fail to produce a PN.
As the AGB-manqu\'e evolution can effectively transfer some fraction
of the pAGB energy budget to the integrated UV flux of the galaxy,
this scenario could also explain the anticorrelation between the
strength of the UV-upturn \citep[as defined by][]{Bur88} and the
luminosity-specific PN number density of early-type galaxies, which
respectively increase and decrease with both galaxy mass and
metallicity.
In fact, it is in M32 that \citet{Bro00} provided some support to this
picture, finding with near-UV Hubble Space Telescope (HST)
observations that the relative faint UV flux of M32 originates from
hot HB stars and their AGB-manqu\'e progeny rather than from their
brighter post-AGB counterparts.
Yet, given the relatively long UV-bright lifetime of stars leaving the
hot end of the HB ($\sim 10^6 - 10^7$yrs compared to $\sim 10^3 -
10^4$ yrs for stars leaving the HB red end) \citet{Bro00} conclude
that only a small fraction of HB stars in M32 live at its blue tip, so
that in fact many more bright pAGB should be observed. The recent
analysis of the UV colour-magnitude diagram (CMD) of M32 by
\citet{Bro08} further reinforces this result.
The findings of \citet{Bro00,Bro08} not only contrast again with the
notion that only hot and less massive HB stars should now form in the
old stellar population of M32 (unless one considers binary
coalescence), but also add an additional problem for the current
models of the late evolution of stars.

\placefigUV

To explain such a dearth of pAGB stars, among other possibilities
\citet{Bro00} suggested that pAGB could be obscured by circumstellar
material, in particular as they cross the UV CMD before reaching the
region where they peak in temperature at around 60,000 K. 
This is an interesting possibility considering that PNe are themselves
affected by dust extinction and given that, assuming a \citet{Car89}
reddening law and R$_{\rm V}=3.1$, the typical extinction of 0.7
magnitude at 5007\AA\ for PNe in both star-forming and old galaxies
\citep{Her09,Ric99,Jac99} would translate into considerable reddening
values of $\sim$1.71 and $\sim$1.67 in the HST far- and near-UV
passbands (FUV and NUV), respectively.
To actually check the impact of reddening in the pAGB population of
M32, or at least for the fraction that is presently powering a PN, we
looked in the catalogue of \citet[][kindly provided by
  T. Brown]{Bro08} for the best candidate central star for our PNe
sources and plotted their position in the UV CMD shown in
Fig.~\ref{fig:UV}. The field-of-view of the \citeauthor{Bro08} HST
observations falls entirely within the area covered by our
\sauron\ data, being $24\arcsec\times24\arcsec$ and only slightly
offset from the center of M32, but unfortunately only 8 of our PNe end
up within it. As we are looking to assess as conservatively as
possible the impact of reddening on pAGB stars, in Fig.~\ref{fig:UV}
we have located not only the closest but also the brightest of the UV
stars near the location of our PNe.
Except for PN 9, we could always find a pretty unique UV star within a
radius of 0\farcs97 (half a FWHM) with a far-UV magnitude that would
stick out from the distribution of magnitude values of all stars
within 2\farcs5 ($3\,\sigma_{\rm PSF}$) of our PN by at least 3 times
the scatter in such values.
Furthermore, except for PNe 12 and 15, such bright UV stars are always
quite close to our PNe, within 0\farcs25. For PN 15 the relative large
distance of the candidate star (0\farcs63) is less of a concern, given
that admittedly our Gaussian model does not fully reproduce the
\Oiii\ flux distribution of this source, but for PN 12 (0\farcs85) it
is possible that the central PN star lies elsewhere near the PN and
therefore also in the CMD of Fig.~\ref{fig:UV}. This is also most
certainly the case of PN 9, for which the plotted position in
Fig.~\ref{fig:UV} should be considered as upper limit for the far-UV
brightness of its central star.
It is important to note, however, that the central stars of even
relatively faint PNe such as PN 9 and 12 should still be intrinsically
fairly bright pAGB or early-post AGB stars (epAGB) and, in particular,
that our PNe could not be powered by stars that have already reached
the white-dwarf (WD) cooling curve. Indeed, the models of \citet[][see
  their Fig.~10]{Mar04} shows that at that stage it is possible to
obtain only PNe of $M_{5007} > 0.5$, which are much fainter than the
PNe that we can detect in M32.
On the other hand, even considering the uncertain position of PN 9 and
possibly PN 12 in Fig.~\ref{fig:UV}, we note that accounting for the
typical reddening observed in PNe would bring back up and left the
position of the central PN candidate stars of our brightest and
faintest PNe near to the location in the UV CMD where pAGB and epAGB
reach their peak temperature and spend most of their time, for FUV
magnitudes around $\sim20$ and $\sim22$, respectively, and a FUV-NUV
colour $\sim-1.5$.
In fact, we note that PNe as bright as our PN 1, 5 and 6, with
$M_{5007} < -3.5$, can only be powered by central stars that came from
main-sequence stars of masses between 1.6 and 2.5 $\rm M_{\odot}$ and
which would cross the UV CMD of Fig.~\ref{fig:UV} along the brightest
pAGB tracks \citep[again, see][]{Mar04}, which implies that these
sources actually {\it must\/} be reddened.

We can double-check this result by using our integral-field data to
estimate the position in the theoretical $\log{T_{\rm eff}} - \log{L}$
Hertzsprung-Russell (HR) diagram of the PNe for which we can also
detect the \Hb\ emission, and then compare their location to the
evolutionary tracks of pAGB stars.
Through the models of \citet[][]{Dop92} it is indeed possible to
estimate the temperature $T_{\rm eff}$ and luminosity $L$ of the PNe
central star using the value for the \OiiioHb\ ratio and the
\Hb\ flux, respectively.
These models predict a conversion efficiency $\epsilon$ between the
bolometric luminosity $L$ of the central star and the
\Hb\ recombination flux from the PN that is rather constant and
independent of the gas temperature $T_{\rm e}$ or metallicity $Z_{\rm
  gas}$. On the other hand, since Oxygen is the main coolant of PNe,
the \Oiii$\lambda5007$ flux depends eavily on $T_{\rm e}$ and $Z_{\rm
  gas}$ and some knowledge of these quantities further helps
constraining $T_{\rm eff}$ with the \OiiioHb\ ratio.
The restricted wavelength range of our \sauron\ data does not grant us
access to the weak \Oiii$\lambda4363$ line that is normally used to
measure $T_{\rm e}$, but we can use our spectroscopic data to measure
the stellar metallicity along the line-of-sight of our PNe and thus
gauge the metallicity of their central star, which in turn relates
naturally to $Z_{\rm gas}$.

There are 7 PNe in the central regions of M32 for which we could
measure the \OiiioHb\ ratio (Tab.~\ref{tab:PNeProp}), with values
ranging from to just above 6 to almost 17. The \sauron\ aperture
spectra around these objects (extracted within a radius
$r=3\,\sigma_{\rm PSF}$, Fig.~\ref{fig:FluxPNeSpectra}) indicate
values for the H$\beta$ absorption line and the [MgFe50]' index of
\citet{Kun10} between 1.77 -- 1.92 \AA\ and 3.01 -- 3.08 \AA,
respectively, which in turn yield an estimate for the stellar
metallicity [Z/H] between -0.4 and -0.33 \citep[see Fig.~3
  of][]{Kun10}, depending on whether one adopts the models of
\citet{Schi07} or of \citet{Tho03}.
With this gauge for the central star metallicity we can infer from
Fig.~3 of \citeauthor{Dop92} a star temperature $T_{\rm eff}$ around
$\sim$55,000 K for the PNe with low \OiiioHb\ values $\sim$ 6 -- 8,
and above $\sim$85,000 K for the PNe with \OiiioHb$ >11$.
From Fig.~1 of Dopita et al. this range of temperatures would then
correspond to an average conversion efficiency $\epsilon$ around
0.7\%, which translate our observed \Hb\ fluxes into central star
luminosities in the $\log(L/L{_\odot})=$3.0 -- 3.2 range.
These $T_{\rm eff}$ and $L$ values would place the 7 PNe with
\OiiioHb\ measurements from our \sauron\ observations right below and
to the right of the post-AGB evolutionary paths that are normaly used
in the HR diagram to trace the luminosity evolution of PNe (see, e.g.,
Fig.~1 of Mendez et al. 2008 and Fig.~3 of Schonberner et al. 2010),
too cold to be on the cooling track of white dwarves (which cannot
power PNe) and too faint to be on par with the pAGB stars that
normally produce PNe (which range in luminosity between
$\log(L/L{\odot})=$3.5 -- 4.0).
Assuming a Solar or nearly Solar metallicity for the central stellar
population of M32, as found by \citet{Ros05}, would lead to an even
more stringent result since in this case the central star $T_{\rm
  eff}$ of the highly-excited PNe would be better constrained to lower
values by the models of Dopita et al.
Thus, also for this second subsample of PNe it would seem necessary to
invoke dust reddening in order to to reconcile our data with the model
expectations, further supporting the possible role of circumstellar
dust in producing the apparent dearth of pAGB stars in M32.

%
\section{Conclusions and Future Prospects}
\label{sec:Conclusions}

Using \sauron\ data for the nearby elliptical M32 we have shown how,
by means of a careful subtraction of the stellar background,
integral-field spectroscopy (IFS) allows to detect the emission of
single Planetary Nebulae (PNe) in the optical regions of early-type
galaxies down to flux levels otherwise hardly accessible to standard
narrow-band photometry.
In turn, this makes it possible to trace the PNe luminosity function
(PNLF) within a wider completeness limit and in galactic regions where
stellar populations gradients can be the most extreme, and therefore
where studying how the shape and normalization of the PNLF originate
from the bulk of a galactic stellar population could be the most
instructive.
Moreover, we have shown that the incompleteness of a survey for PNe in
the optical region of a galaxy can be very well understood when using
IFS data, and that the possibility to extract also the stellar
kinematics further allows to check the membership of PNe on a local
basis, that is, along the line-of-sight towards a given PN.
Such a local comparison between the PNe and stellar kinematics could
be useful to identify subpopulations of PNe of different origins.
Finally, although with the \sauron\ data we could only confirm the
high-ionisation nature of nearly half of our sources, on a longer
wavelength range IFS has a great potential to study the nebular
spectrum of extragalactic PNe, as already illustrated in the
pioneering work of \citet{Rot04}.

In the specific case of M32, with just two 10-minutes-long
\sauron\ pointings, we have doubled the number of known PNe within the
effective radius of this galaxy and detected PNe five times fainter
than previously found with narrow-band imaging by \citet{Cia89}, while
collecting slightly less the same number of photons.
Furthermore, accounting for the incompleteness of our survey across
the entire luminosity range spanned by our detected PNe, we have
concluded that the central PNe population of M32 is consistent with
the generally adopted shape for the PNLF (that of Ciardullo et al.)
and its typical normalization observed in early-type galaxies
(i.e. with a PNe number density $\alpha$ of a few $10^{-7}$ per ${\rm
  L_{\odot}}$).
Finally, thanks to exquisite \HST\ data and to the proximity of M32 we
were able to combine our PNe measurements with images for the resolved
UV-bright stellar population of M32 and conclude that the PNe central
stars may be considerably reddened by dust, a result that we could
double-check thanks to the IFS nature of our data by estimating the
temperature and luminosity for the central star of the PNe with
\OiiioHb\ measurements.
This finding would support the suggestion by \citeauthor{Bro08} that
circumnstellar dust extinction could explain the apparent dearth of
pAGB stars in M32, a problem that may also apply to the halo of the
Milky Way \citep{Wes10}.

Given the intrinsically small number of PNe that is expected in the
central regions of such a small galaxy as M32, it is unlikely that
with the current instruments (e.g., \sauron\ or \Vimos) further IFS
observations will reveal a sufficiently larger number of objects to
add either to what found in this work or to what is already known in
the outskirts of M32 from narrow-band imaging or slit-less
spectroscopy, even though more IFS data could allow for more
emission-line diagnostics.
 
On the other hand, there is a wealth of \sauron\ data for many more
early-type galaxies that was aquired over the course of the
\sauron\ \citep{deZ02} and ATLAS$^{\rm3D}$ \citep{Cap11} surveys where
the presence of PNe have not yet been systematically investigated.
In fact, even though a few PNe were detected in \citet{Sar06}, many
more PNe have certainly been missed because a large fraction of the
data were spatially binned in order to extract a reliable stellar
kinematics and due to the common presence of diffuse ionised
gas. Indeed spatial binning washes out the signal of the weaker PNe
against the stellar background, whereas extended nebular emission
makes it difficult to detect PNe unless their distinctive spectrum
characterised by high-ionisation and narrow lines is purposely looked
for.
Paying attention also to the unresolved nature of the PNe emission and
working with unbinned data using an educated guess for the stellar
kinematics \citep[for instance based on relatively simple but
  functional dynamical models, see][]{Cap08} it may be possible to tap
a significant PNe population in the central regions of nearby
early-type galaxies. Indeed, although the \sauron\ and ATLAS$^{\rm3D}$
surveys have targeted much more distant objects than M32, the ability
to carefully subtract the stellar background will still enable the
detection of fainter PNe in the central regions of these galaxies
compared to narrow-band or slit-less spectroscopic
surveys. 
Furthermore, the relatively large number of objects surveyed by these
campaigns (260) will allow to combine the measurements from different
galaxies, and for instance to better explore the bright end of the
PNLF or variations of its shape and normalization as a function of the
stellar age and metallicity that will also be provided by the these
\sauron\ data.

To conclude, given the number of findings obtained here with a modest
time investment (not even 30 minutes with overheads) on a 4m
telescope, it is exciting to consider what could be learned on the PNe
population in the optical regions of other galaxies, when the next
generation of integral-field spectrographs will be mounted on 8m-class
telescopes, such as MUSE on VLT \citep[see, e.g.,][]{Bac06}.
For instance, with a longer wavelength range and a larger light
bucket, MUSE will not only find many more and fainter PNe, but it will
also allow full diagnostic of their nebular spectrum that could lead
to a direct measurement of the gas metallicity and to recognise other
nebular sources that could pose as a PNe in narrow-band surveys
\citep[see, e.g.,][]{Fre10}. In turn the PNe metallicity could be
compared with that of the underlying stellar population, which would
also be better constrained thanks to an extended spectral range.
Furthermore, the spatial resolution that will be provided to MUSE by
the use of adaptive optics will allow further comparisons with
\HST\ or other space-based measurements for the resolved stellar
population of nearby galaxies, such as those presented here for M32,
which are bound to shed more light on the link between PNe and their
parent stellar population.

\section*{Acknowledgements}
MS is grateful to Thomas Brown, Paola Marigo, Harald Kuntschner, Ralph
Napiwotzki and Brent Miszalski for the useful discussions, and to
Thomas Brown for also providing the near- and far-UV HST data of
M32. We are also indebted to the referee, Robin Ciardullo, for his
many suggestions. MS also acknowledges support from his STFC Advanced
Fellowship (ST/F009186/1) and the Institut d'Astrophysique de Paris
for its hospitality during part of the preparation of this paper,
whereas MC acknowledges support from a STFC Advanced Fellowship
(PP/D005574/1) and a Royal Society University Research Fellowship.

%

\label{lastpage}

\begin{thebibliography}{}
\bibitem[Bacon et al.(2001)]{Bac01}
Bacon R., et al., 2001, MNRAS, 326, 23

\bibitem[Bacon et al.(2006)]{Bac06} 
Bacon R., et al., 2006, The Messenger, 124, 5 

\bibitem[\protect\citeauthoryear{Brown et al.}{2000}]{Bro00} 
Brown T.~M., Bowers C.~W., Kimble R.~A., Sweigart A.~V., Ferguson
H.~C., 2000, ApJ, 532, 308

\bibitem[\protect\citeauthoryear{Brown et al.}{2008}]{Bro08} 
Brown T.~M., Smith E., Ferguson H.~C., Sweigart A.~V., Kimble R.~A.,
Bowers C.~W., 2008, ApJ, 682, 319

\bibitem[\protect\citeauthoryear{Bruzual \& Charlot}{2003}]{Bru03} 
Bruzual G., Charlot S., 2003, MNRAS, 344, 1000

\bibitem[\protect\citeauthoryear{Burnstein et al.}{1988}]{Bur88}
Burstein, D., Bertola, F, Buson, L.~M., Faber, S.~M., Lauer, T.~R.,
1988, ApJ, 328, 440

\bibitem[\protect\citeauthoryear{Buzzoni, Arnaboldi, \& Corradi}{2006}]{Buz06}
Buzzoni A., Arnaboldi M., Corradi R.~L.~M., 2006, MNRAS, 368, 877

\bibitem[\protect\citeauthoryear{Cardelli, Clayton, \& Mathis}{1989}]{Car89}
Cardelli J.~A., Clayton G.~C., Mathis J.~S., 1989, ApJ, 345, 245 

\bibitem[Cappellari \& Copin(2003)]{Cap03} Cappellari M., Copin Y.,
2003, \mnras, 342, 345

\bibitem[Cappellari \& Emsellem(2004)]{Cap04}
Cappellari M., Emsellem E., 2004, PASP, 116, 138

\bibitem[\protect\citeauthoryear{Cappellari et al.}{2006}]{Cap06}
Cappellari M., et al., 2006, MNRAS, 366, 1126

\bibitem[\protect\citeauthoryear{Cappellari}{2008}]{Cap08} 
Cappellari M., 2008, MNRAS, 390, 71 

\bibitem[\protect\citeauthoryear{Cappellari et al.}{2011}]{Cap11}
Cappellari M., et al., 2011, MNRAS, in press (doi:10.1111/j.1365-2966.2010.18174.x)

\bibitem[\protect\citeauthoryear{Ciardullo et al.}{1989}]{Cia89}
Ciardullo R., Jacoby G.~H., Ford H.~C., Neill J.~D., 1989, ApJ, 339, 53 

\bibitem[\protect\citeauthoryear{Ciardullo et al.}{2004}]{Cia04}
Ciardullo R., Durrell P.~R., Laychak M.~B., Herrmann K.~A., Moody K.,
Jacoby G.~H., Feldmeier J.~J., 2004, ApJ, 614, 167

\bibitem[\protect\citeauthoryear{Ciardullo et al.}{2005}]{Cia05} 
Ciardullo R., Sigurdsson S., Feldmeier J.~J., Jacoby G.~H., 2005, ApJ,
629, 499

\bibitem[\protect\citeauthoryear{Ciardullo}{2006}]{Cia06}
Ciardullo R., 2006, IAUS, 234, 325

\bibitem[\protect\citeauthoryear{de Zeeuw et  al.}{2002}]{deZ02} 
de Zeeuw P.~T., et al., 2002, MNRAS, 329, 513

\bibitem[\protect\citeauthoryear{Dopita, Jacoby, \& Vassiliadis}{1992}]{Dop92}
Dopita M.~A., Jacoby G.~H., Vassiliadis E., 1992, ApJ, 389, 27 

\bibitem[\protect\citeauthoryear{Dopita et al.}{1997}]{Dop97} 
Dopita M.~A., et al., 1997, ApJ, 474, 188 

\bibitem[\protect\citeauthoryear{Douglas et al.}{2007}]{Dou07}
Douglas N.~G., et al., 2007, ApJ, 664, 257

\bibitem[Emsellem et al.(2004)]{Ems04}
Emsellem, E., et al.\  2004, MNRAS, 352, 721

\bibitem[Falc{\' o}n-Barroso et al.(2006)]{Fal06}
Falc{\' o}n-Barroso, J., et al.\ 2006, MNRAS, 369, 529

\bibitem[\protect\citeauthoryear{Frew \& Parker}{2010}]{Fre10}
Frew D.~J., Parker Q.~A., 2010, PASA, 27, 129 

\bibitem[\protect\citeauthoryear{Jacoby, Ciardullo, \& Ford}{1990}]{Jac90}
Jacoby G.~H., Ciardullo R., Ford H.~C., 1990, ApJ, 356, 332 

\bibitem[\protect\citeauthoryear{Jacoby et al.}{1992}]{Jac92} 
Jacoby G.~H., et al., 1992, PASP, 104, 599

\bibitem[\protect\citeauthoryear{Jacoby \& Ciardullo}{1999}]{Jac99}
Jacoby G.~H., Ciardullo R., 1999, ApJ, 515, 169 

\bibitem[\protect\citeauthoryear{Henize \& Westerlund}{1963}]{Hen63}
Henize K.~G., Westerlund B.~E., 1963, ApJ, 137, 747

\bibitem[\protect\citeauthoryear{Herrmann \& Ciardullo}{2009}]{Her09}
Herrmann K.~A., Ciardullo R., 2009, ApJ, 703, 894

\bibitem[\protect\citeauthoryear{Lauer et al.}{1998}]{Lau98}
Lauer T.~R., Faber S.~M., Ajhar E.~A., Grillmair C.~J., Scowen P.~A.,
1998, AJ, 116, 2263

\bibitem[\protect\citeauthoryear{Kuntschner et al.}{2010}]{Kun10}
Kuntschner H., et al., 2010, MNRAS, 408, 97 

\bibitem[\protect\citeauthoryear{Mamon, Dekel, \& Stoehr}{2005}]{Mam05} Mamon G.~A., Dekel A., Stoehr F., 2005, AIPC, 804, 345 

\bibitem[\protect\citeauthoryear{Marigo et al.}{2004}]{Mar04}
Marigo P., Girardi L., Weiss A., Groenewegen M.~A.~T., Chiosi C.,
2004, A\&A, 423, 995

\bibitem[\protect\citeauthoryear{M{\'e}ndez et al.}{2008}]{Men08}
M{\'e}ndez R.~H., Teodorescu A.~M., Sch{\"o}nberner D., Jacob R., Steffen M., 2008, ApJ, 681, 325 

\bibitem[\protect\citeauthoryear{Merrett et al.}{2006}]{Mer06}
Merrett H.~R., et al., 2006, MNRAS, 369, 120

\bibitem[\protect\citeauthoryear{Peletier}{1993}]{Pel93}
Peletier R.~F., 1993, A\&A, 271, 51

\bibitem[\protect\citeauthoryear{Richer, Stasi\'nka, \& McCall}{1999}]{Ric99}
Richer M.~G., Stasi\'nka G., McCall M.~L.~M., 1999, A\&AS, 135, 205

\bibitem[\protect\citeauthoryear{Romanowsky et al.}{2003}]{Rom03} 
Romanowsky A.~J., Douglas N.~G., Arnaboldi M., Kuijken K., Merrifield
M.~R., Napolitano N.~R., Capaccioli M., Freeman K.~C., 2003, Sci, 301,
1696

\bibitem[\protect\citeauthoryear{Rose et al.}{2005}]{Ros05}
Rose J.~A., Arimoto N., Caldwell N., Schiavon R.~P., Vazdekis A.,
Yamada Y., 2005, AJ, 129, 712

\bibitem[\protect\citeauthoryear{Roth et al.}{2004}]{Rot04} 
Roth M.~M., Becker T., Kelz A., Schmoll J., 2004, ApJ, 603, 531

\bibitem[\protect\citeauthoryear{Sambhus, Gerhard \& Mendez}{2006}]{Sam06}
Sambhus N., Gerhard, O., \& Mendez, R.~H., 2006, AJ, 131, 837

\bibitem[\protect\citeauthoryear{S{\'a}nchez-Bl{\'a}zquez et al.}{2006}]{San06}
S{\'a}nchez-Bl{\'a}zquez P., et al., 2006, MNRAS, 371, 703

\bibitem[\protect\citeauthoryear{Sarazin, Irwin \& Bregman}{2000}]{Sar00}
Sarazin C.~L., Irwin, J.~A., \& Bregman, J.~N., 2000, ApJ, 544, L101

\bibitem[\protect\citeauthoryear{Sarzi et al.}{2006}]{Sar06}
Sarzi M., et al., 2006, MNRAS, 366, 1151

\bibitem[\protect\citeauthoryear{Sarzi et al.}{2010}]{Sar10}
Sarzi M., et al., 2010, MNRAS, 402, 2187

\bibitem[\protect\citeauthoryear{Schiavon}{2007}]{Schi07} 
Schiavon R.~P., 2007, ApJS, 171, 146

\bibitem[\protect\citeauthoryear{Sch{\"o}nberner et al.}{2007}]{Sch07}
Sch{\"o}nberner D., Jacob R., Steffen M., Sandin C., 2007, A\&A, 473, 467 

\bibitem[\protect\citeauthoryear{Sch{\"o}nberner et al.}{2010}]{Sch10}
Sch{\"o}nberner D., Jacob R., Sandin C., Steffen M., 2010, A\&A, 523, A86 

\bibitem[\protect\citeauthoryear{Thomas, Maraston, \& Bender}{2003}]{Tho03}
Thomas D., Maraston C., Bender R., 2003, MNRAS, 339, 897 

\bibitem[\protect\citeauthoryear{Tonry et al.}{2001}]{Ton01} 
Tonry J.~L., Dressler A., Blakeslee J.~P., Ajhar E.~A., Fletcher
A.~B., Luppino G.~A., Metzger M.~R., Moore C.~B., 2001, ApJ, 546, 681

\bibitem[\protect\citeauthoryear{Weston, Napiwotzki, \& Catal{\'a}n}{2010}]{Wes10}
Weston S., Napiwotzki R., Catal{\'a}n S., 2010, AIPC, 1273, 197 


\bibitem[\protect\citeauthoryear{White, Sarazin \& Kulkarni}{2002}]{Whi02}
White, R.~E., Sarazin C.~L., \& Kulkarni, S.~R., 2002, ApJ, 571, L23


\end{thebibliography}
\end{document}